\documentclass[fleqn,usenatbib]{mnras}
\usepackage[utf8]{inputenc}
\usepackage{newtxtext,newtxmath}
\usepackage{comment}

\usepackage{soul}
\usepackage[T1]{fontenc}

\DeclareRobustCommand{\VAN}[3]{#2}
\let\VANthebibliography\thebibliography
\def\thebibliography{\DeclareRobustCommand{\VAN}[3]{##3}\VANthebibliography}

\usepackage{graphicx}	
\usepackage{amsmath}	
\usepackage{amsmath}
\usepackage{geometry}
\usepackage{caption}
\usepackage{hyperref}
\usepackage{lscape}
\usepackage{amsfonts}

\def \inte {{\it INTEGRAL}}
\def \swift {{\it Swift}}

\def \xmm {{\it XMM}-Newton}
\def \src{{SRGA J144459.2--604207}}
\def \obj{{SRGA J1444}}
\def \swiftxrt{{\it Swift}-XRT}

\def \nustar{{\it NuSTAR}}

\def \nicer{{\it NICER}}

\def \ixpe{{\it IXPE}}

\def \relxill{{\tt relxill}}

\def \ninjasat{{\it NinjaSat}}

\newcommand{\erg}{erg cm$^{-2}$ s$^{-1}$} 
\newcommand{\lum}{erg s$^{-1}$}

\title[A broadband view at \src{}]
{Relativistic X-ray reflection and thermonuclear burst from accreting millisecond X-ray pulsar \src{}}

\author[Mandal et al.]{
Manoj Mandal\,$^{1}$\thanks{E-mail: manojmandal213@gmail.com},
Sachindra Naik\,$^{1}$,
Birendra Chhotaray$^{1}$ \\
$^{1}$Astronomy and Astrophysics Division, Physical Research Laboratory, Navrangpura, Ahmedabad - 380009, Gujarat, India
}
\date{Accepted XXX. Received YYY; in original form ZZZ}

\pubyear{2015}

\begin{document}
\label{firstpage}
\pagerange{\pageref{firstpage}--\pageref{lastpage}}
\maketitle

\begin{abstract}
We present the results obtained from the spectral and temporal study of thermonuclear bursts from the millisecond X-ray pulsar \src{} detected with \nicer{}. The dynamic evolution of the spectral parameters in a broad energy range is also investigated during a simultaneously detected burst with \xmm{} and \nustar{}. The burst profiles exhibit a strong energy dependence, as observed with \xmm{}, \nicer{}, and \nustar{}. We investigated the reflection feature during these bursts using the disk reflection model. As observed during the peak of the \nicer{} bursts, the reflection model can contribute $\sim$30\% of the overall emission. During the \nicer{} bursts, a correlation is observed between the flux of the blackbody and the reflection components. The measurements of the mass accretion rate indicate that the bursts may be powered by a mixed H/He fuel. Moreover, the broadband \nicer{} and \nustar{} spectra are also used to probe the reflection signature in the burst-free persistent region using the relativistic reflection model. Based on the variability of the count rate during the \nustar{} observation, we also investigate the evolution of spectral parameters during two different flux levels of the \nustar{} observation. The inner disk radius (R$_{\textrm{in}}$) and the angle of inclination are found to be nearly 11 $R_{g}$ and $\sim50^{\circ}$, respectively. The magnetic field strength at the poles of the neutron star is estimated to be $ B~\sim6.0 \times 10^8$ G, assuming that the inner disk is truncated at the magnetospheric boundary. 
\end{abstract}



\section{Introduction}
\label{intro}
 A neutron star low-mass X-ray binary (LMXB) system consists of a neutron star (NS) and a low-mass companion star ($M_d\leq 1\,M_\odot$) orbiting around the common center of mass. In these LMXBs, the compact object accretes matter from the low-mass companion through the Roche-lobe overflow mechanism, forming an accretion disk.
 In the case of high magnetic field ($\sim$10$^{12}$ G; \citealt{Staubert2019}) neutron star LMXBs, the accretion disk gets truncated at the magnetospheric radius which is relatively large compared to that in the case of low magnetic field ($\sim$10$^{7-9}$ G; \citealt{Ca09,Mu15}) neutron star LMXBs. In the case of neutron star LMXBs with a low magnetic field, the accretion disk may get extended close to the NS surface. In this case, the accreted material may directly accumulate onto the NS surface, unlike in the case of high magnetic field neutron star LMXBs, where accretion takes place at the magnetic poles of the neutron star. In the low-magnetic field neutron star LMXBs, the accreted material consisting of hydrogen and/or helium burns unstably on the surface of the NS, resulting in thermonuclear X-ray bursts \citep{Le93, Ga06}. 

  Accreting millisecond pulsars (AMXPs) are a subclass of NS LMXBs, showing coherent pulsation of a few milliseconds (for review \citealt{Po06}). These systems display thermonuclear X-ray bursts, which can be utilized to understand accretion physics and the burning mechanism \citep{Salvo2020, Alessandro2021}. The AMXPs are rotating at a spin frequency in the range of 180--600 Hz. The AMXPs are considered to be a unique laboratory for studying the thermonuclear X-ray burst and its effect on the truncated accretion disk at the magnetosphere boundary. During thermonuclear X-ray bursts, the intensity increases rapidly as a result of unstable burning on the NS surface \citep{Ga08}. The surface emission is often more than ten times the persistent X-ray emission during a thermonuclear burst \citep{Ba10}. Thermonuclear X-ray bursts are usually short, with a rapid rise (0.5--5 s) and exponential decay (10--100 s) \citep{Le93}. The lighter species are transformed into heavier elements during the burst through nuclear chain reactions \citep{Le93, St03, Sc06}. Sometimes, the peak luminosity during the burst is high enough to lift the photosphere of the NS, resulting in a photosphere radius expansion (PRE) \citep{Le93, Ta84}. The photosphere reaches its maximum expansion during PRE, and then the blackbody temperature decreases at a constant luminosity level. As the process comes to an end, the photosphere approaches the NS surface. The burst spectrum can be described by an absorbed blackbody model with a temperature of $\sim$0.5--3 keV, assuming that the NS emits like a blackbody \citep{Va78, Ku03}.  The accretion rate sometimes increased due to the Poynting-Robertson drag, resulting in an enhanced persistent emission \citep{Walker1992, Worpel2013}. In this case, the burst spectrum deviates from pure blackbody emission, and a variable persistent emission modeling approach ($f_a$) is used to model the spectrum. As an observational consequence, a soft and a hard excess are found in the burst spectrum \citep{Worpel2013}. The excess can be explained as the reflected burst emission from the accretion disk \citep{Lu23, Zh22}. Sometimes, emission lines and edges are also observed during a super burst (e.g, IGR~J17062-6143, 4U~1820-30), which can be alternatively described by the reflection from the photoionized disk \citep{Ke17, Ba04}. The high-energy X-ray photons during the burst may interact with the hot corona, the NS atmosphere, and the surrounding accretion disk, leading to several observable consequences \citep{De18}.  

 The X-ray reflection features have been observed in several NS LMXBs, such as Ser X-1 \citep{Bh07, Ca08}, 4U 1820–30 \citep{Ca08}, GX 349+2 \citep{Ca08}, Aql X-1 \citep{Ma25}, and 4U 1702-429 \citep{Lu19, 4U1702}. Especially in accreting millisecond pulsars (AMXPs) or bursting sources, the reflection spectral modeling approach is a powerful tool for investigating disk properties, assuming that the disk is truncated by the rapidly rotating magnetosphere close to the neutron star surface. This can also be used to constrain the magnetic field of the neutron star \citep{Ca09, Lu19}. The outer layers of the accretion disk likewise reprocess the irradiating hard X-ray photons and emit reprocessed emission, regardless of where the thermal X-ray photons come from. The energy spectrum associated with the reprocessed photons exhibit several complex features, such as an iron (Fe~K$\alpha$) emission line at $\sim$6.4 keV, and a Compton back-scattering hump in the hard X-ray energy band at $\sim$20 keV superimposed on a continuum known as the ``reflection'' spectrum \citep{Ge91, Fa89, Fa00, Mi07}. Scattering of the X-ray photons with the heated inner flow, the Doppler effect, and general relativistic effects all contribute to the broadening of the Fe~K$\alpha$ emission line \citep{Fa89, Fa00, Mi07}.

The accreting millisecond pulsar \src{} (\obj{} hereafter) was discovered in 2024 February \citep{Me24} with the Mikhail Pavlinsky ART-XC telescope on the Spectrum-Roentgen-Gamma (SRG) observatory \citep{Pa21, Su21}. Following the discovery, the source was observed in multiple wavelength bands. \citet{Ng24} reported a thermonuclear burst and discovered coherent X-ray pulsations at $\sim$447.9 Hz with a orbital period of $\sim$5.2 hr with \nicer{}. Thermonuclear bursts are also reported using the \inte{} \citep{Sa24a, Sa24b}, \ninjasat{} \citep{Ta24}, \xmm{}, and \nustar{} observations \citep{Malacaria2025}. The \ixpe{} revealed polarized emission from \obj{} with an average polarization degree of $\sim$2.3\% \citep{Pa25}.

In this work, we report the results of the detailed time-resolved spectral study of newly found thermonuclear bursts using \nicer{}. We also analyzed bursts detected simultaneously in the \xmm{} and \nustar{} observations. The energy dependence of the burst profiles is studied for the bursts detected with \nicer{}, \xmm{}, and \nustar{}. We also investigated the reflection features during bursts to probe the burst-disk interaction using the disk reflection model {\tt relxillNS}. We explored the reflection features from the persistent emission using the self-consistent relativistic reflection modeling approach to investigate disk properties and geometry. The \nicer{}, \xmm{}, and \nustar{} observations are used to examine the reflection feature during thermonuclear bursts as well as in the burst-free persistent emission regions. Using the results, we derived disk parameters, geometry, the magnetic dipole moment, and the magnetic field for the accreting millisecond pulsar. The paper is organized as follows. The observations and data analysis methodology are described in Section~\ref{obs}. Section~\ref{res} summarizes the results of the spectral and temporal study. The discussion of the results is presented in Section~\ref{dis}. The summary and conclusions are presented in Section~\ref{con}.

 \begin{figure}
\centering{
\includegraphics[width=8.0 cm]{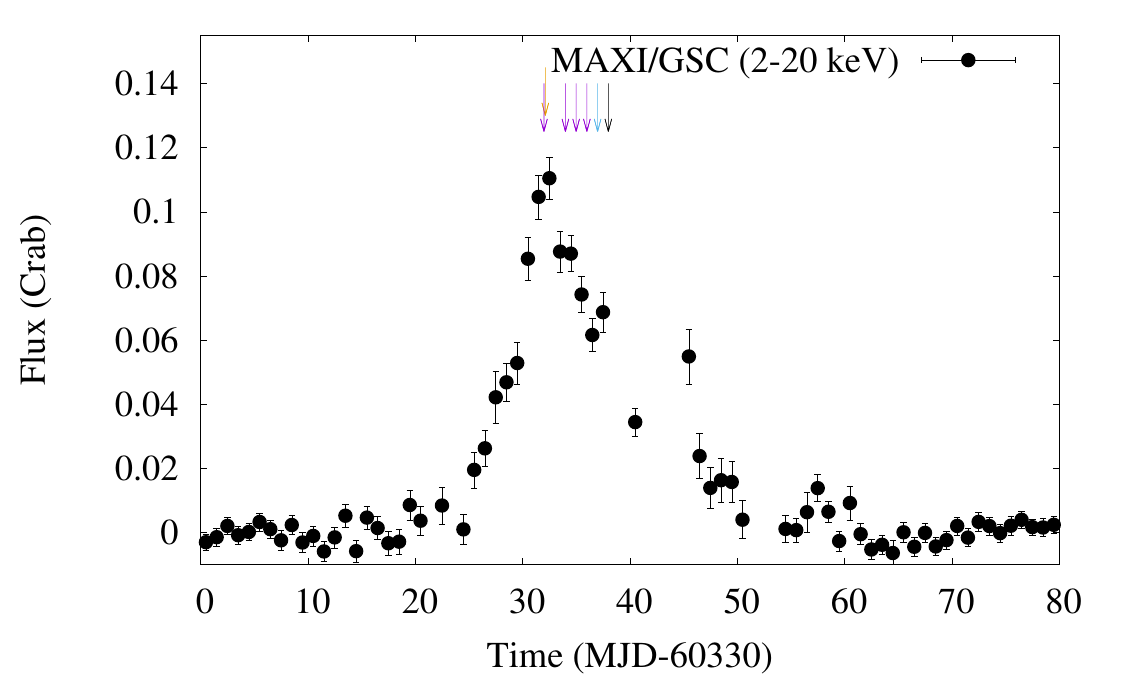}
\caption{{MAXI/GSC daily light curve of \obj{} during the 2024 outburst. The \swiftxrt{}, \nicer{}, \nustar{}, and \xmm{} observation epochs are shown with the yellow, purple, blue, and black arrows, respectively.} 
}
\label{fig:MAXI}}
\end{figure}
 \begin{figure}
\centering{
\includegraphics[width=8cm]{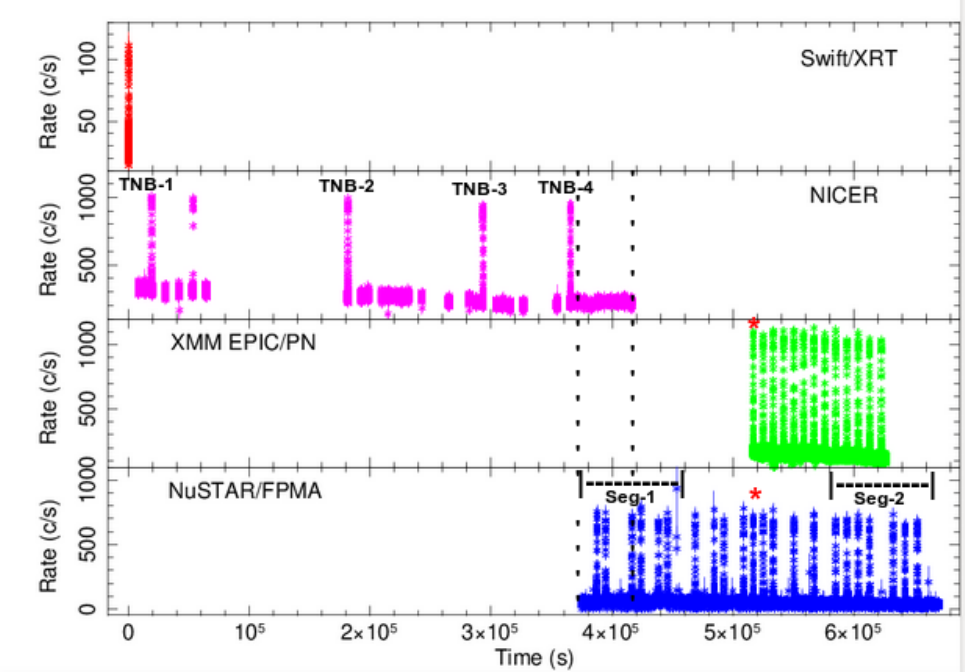}
\caption{The 1 s binned light curves of \obj{} with \nustar{}, \nicer{}, \swiftxrt{}, and \xmm{}. Time-resolved spectral study was performed for one of the bursts simultaneously detected with \xmm{} and \nustar{} (marked with red asterisks). The dotted horizontal lines in the bottom panel indicate the two segments at different flux levels, which are used to investigate the evolution of spectral parameters. Simultaneous \nicer{} and \nustar{} data between the two vertical dotted lines are used for broadband spectral study.}
\label{fig:burst_combined}}
\end{figure}
\begin{table*}
\centering
\caption{Summary of observations and X-ray bursts detected from \obj{} using \nustar{}, \swiftxrt{}, \xmm{}, and \nicer{} observations. The peak flux mentioned for \nustar{} is estimated from the simultaneous spectral fitting of burst spectra from \nustar{} and \xmm{} observations. The reported unabsorbed peak flux is estimated in the 0.1--100 keV range (in units of 10$^{-8}$ erg cm$^{-2}$ s$^{-1}$).}
\begin{tabular}{|l|c|c|c|c|c|} 
\hline	
Observatories	 & Obs. T$_\textrm{start}$     & No. of  & Observation ID & Exposure   & Peak flux\\
		   &   (MJD)       & bursts   &       & (ks) &                \\
\hline
{\it Swift}& 60362.12   &   1 & 00016537002 & 0.5     &  \\
\hline
{\it NICER}& 60362.03  &   2 & 6204190102 (obs-1) & 4.5         & $3.5 \pm 0.1$ \\
& 60364.22    & 1 & 6639080101 (obs-2) & 5.6     & $3.2\pm0.1$  \\
& 60365.19   &  1 & 6639080102 (obs-3) & 3.8       & $3.1\pm0.1$ \\
& 60366.22   & 1 & 6639080103 (obs-4) & 5.2      & $3.3\pm0.1$  \\
\hline
{\it NuSTAR} &  60366.46  & 23 & 80901307002 & 157    & $2.8\pm0.1$ \\
{\xmm}& 60367.96   & 13    & 0923171501 & 135   &  $2.8\pm0.1$
\\
\hline
\label{tab:log_table_burst}
	\end{tabular}
\end{table*}
\begin{figure*}
\centering{
\includegraphics[width=5.8cm]{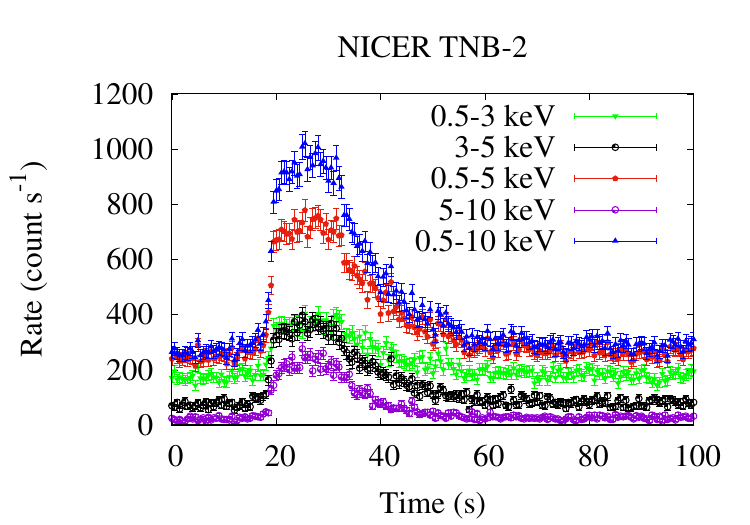}
\includegraphics[width=5.8cm]{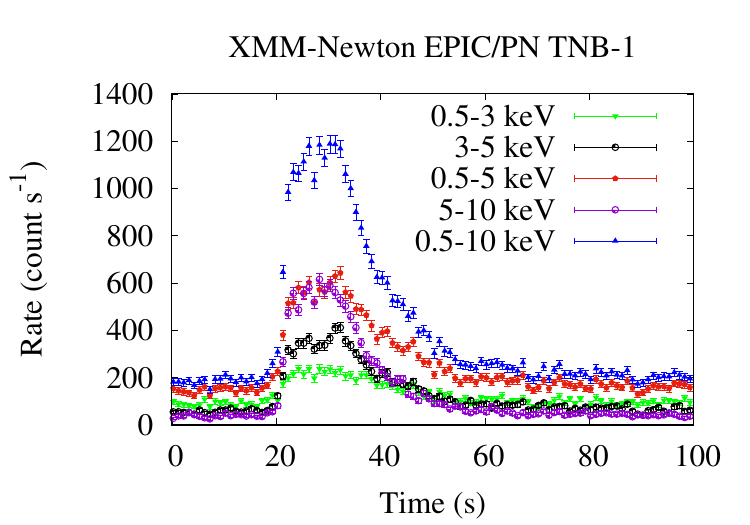}
\includegraphics[width=5.8cm]{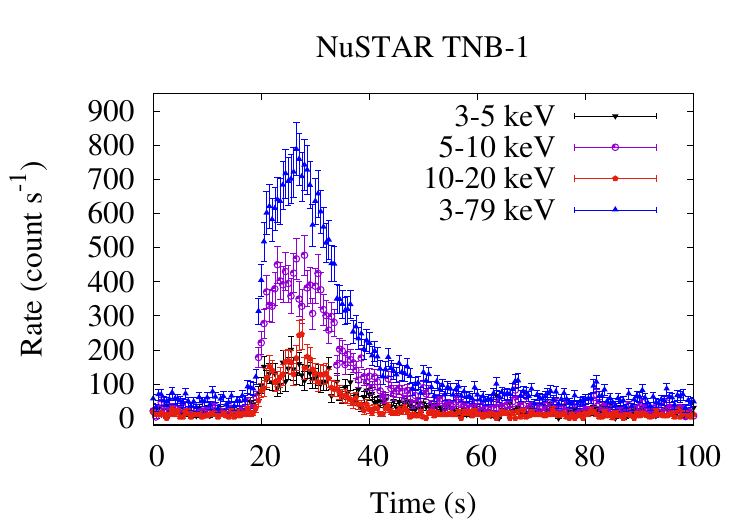}
\caption{Energy-resolved burst profiles of \obj{} are shown using \nicer{}, \xmm{} and \nustar{}.}
	 \label{fig:burst_profile}}
\end{figure*}

\section{Data analysis and methodology}
\label{obs}

In this work, we used publicly available data from \swift{}, \nicer{}, \xmm{}, and \nustar{} observatories during the 2024 outburst of \obj{}. Data are reduced using {\tt HEASOFT} version 6.31.1. In this work, we also used the final data products (light curves) provided by {\it MAXI} \citep{Ma09}.

\subsection{\swift{} observation}

The {\it Swift}/X-ray Telescope (XRT; \citet{Ge04}) observed \obj{} on 22 February 2024. The details of the observation are summarized in Table~\ref{tab:log_table_burst}. We used one \swift{} observation of the source, during which a thermonuclear burst was detected. \obj{} was observed in the Windowed Timing (WT) mode. The \swift{}/XRT data are analyzed using {\tt HEASOFT} version 6.31.1 along with the CALDB version of 20240522. The {\tt xrtpipeline} (version 0.13.7) is used to filter, screen, and reduce the {\it Swift}/XRT data. A circular region of 30 arcsec radius, centered at the source position, is used to extract light curves and spectra. The background spectrum is generated using a circular region of 60 arcsec away from the source position. The {\tt FTOOL XSELECT}(V2.5b) is used to extract the source and background light curves and spectra. The {\tt xrtexpomap} and {\tt xrtmkarf} are used to create the exposure map and ancillary files, respectively. The extracted spectra are used along with the most recent RMF files from CALDB for further spectral analysis.

\subsection{\nicer{} observation}

The Neutron Star Interior Composition Explorer (\nicer{}) onboard the International Space Station is a non-imaging telescope operating in the soft X-ray energy range of 0.2--12 keV \citep{Ge16}. The millisecond X-ray pulsar \obj{} was monitored with \nicer{} during the 2024 outburst. We used four \nicer{} observations during which thermonuclear X-ray bursts were observed. The observation details are summarized in Table~\ref{tab:log_table_burst}. The raw data are processed using the tool {\tt NICERDAS} in {\tt HEASOFT} with the {\tt CALDB} version xti20221001. The cleaned event files are generated from the raw data after applying the standard calibration and filtering tool {\tt nicerl2}. The {\tt barycorr} tool is used to perform the barycentric correction. The {\tt XSELECT} package is utilized to generate light curves and spectra from the cleaned event files. The {\tt nibackgen3C50} tool \citep{Re22} is used to create background spectra corresponding to each \nicer{} observation. The {\tt nicerrmf} and {\tt nicerarf} tools are used to generate response and ancillary response files for the spectral study. The {\tt grppha} task is used to group the spectra for a minimum of 25 counts/bin. 

\subsection{\xmm{} observation}
The European Photon Imaging Camera (EPIC, 0.1-15 keV) is mounted at the focus of each of the three 1500 cm$^2$ X-ray telescopes of the \xmm{} observatory \citep{Ja01}. Two of the imaging spectrometers use MOS CCDs \citep{Tu01}, while one uses PN CCD \citep{St01}. Two identical Reflection Grating Spectrometers (RGS) are placed behind the telescopes and operate in the 0.35-2.5 keV range \citep{He01}. The source was observed with the XMM-Newton observatory on 27 February 2024 for an exposure of $\sim$135 ks.  The source was observed with the EPIC-PN timing mode and also with the RGS. The raw XMM-Newton data are reduced using \xmm{} Science Analysis System ({\tt SAS}) version 20.0.0. The events are filtered using the {\tt evselect} routine with the criteria PATTERN $\le4$ and FLAG = 0. The EPIC-PN  event files are generated using {\tt epproc}, following the {\tt evselect} routine to create source and background light curves and spectra. For the EPIC-PN data analysis, the source events are extracted from a 20-pixel-wide strip centered at the source position RAWX = 37 (i.e., RAWX in [27:47]), while the background region is selected from a source-free region in RAWX columns [5:15] from the same CCD. Spectral response files are generated using the tasks {\tt rmfgen} and {\tt arfgen}. Subsequently, background subtraction and corrections are performed using {\tt epiclccorr}. 

\subsection{\nustar{} observation}

The Nuclear Spectroscopic Telescope Array (\nustar{}) consists of two identical detectors (FPMA/FPMB) that are co-aligned and operate in the 3-79 keV energy range \citep{Ha13}. \nustar{} observed the source on 26 February 2024, for an exposure of $\sim$157 ks. The details of the observation are presented in Table~\ref{tab:log_table_burst}. The standard \nustar{} data analysis software (NUSTARDAS v2.1.2) in {\tt HEASOFT} v 6.31.1, along with the latest {\tt CALDB} version 20240325, is used for data reduction. The {\tt nupipeline} task is used to filter the event files. A circular region of 120 arcsec radius, centered on the source position, is used to extract the source events. A similar circular region, located away from the source, is selected to create background events. The task {\tt nuproducts} is used to generate light curves, spectra, the auxiliary response file, and the response matrix file for both detectors. The light curves are background corrected using the {\tt lcmath} task. 

\begin{figure*}
    \includegraphics[width=0.65\columnwidth]{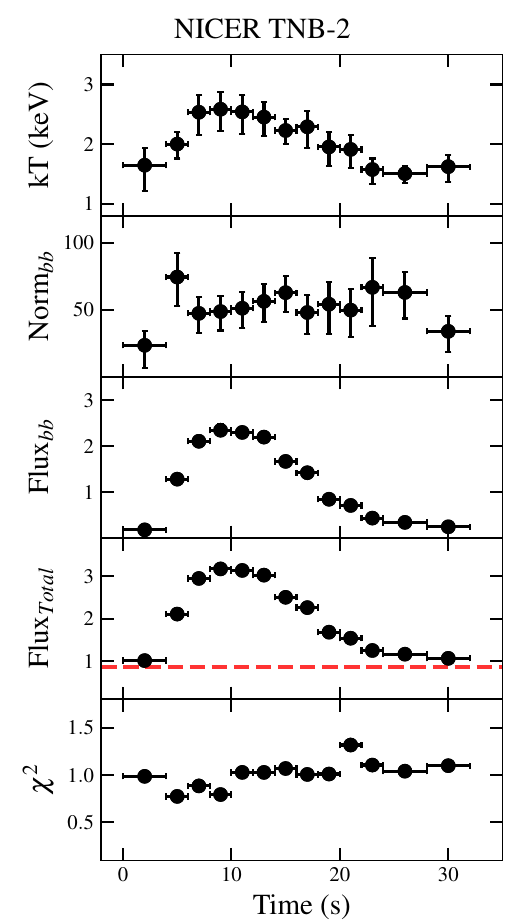} 
       \includegraphics[width=0.65\columnwidth]{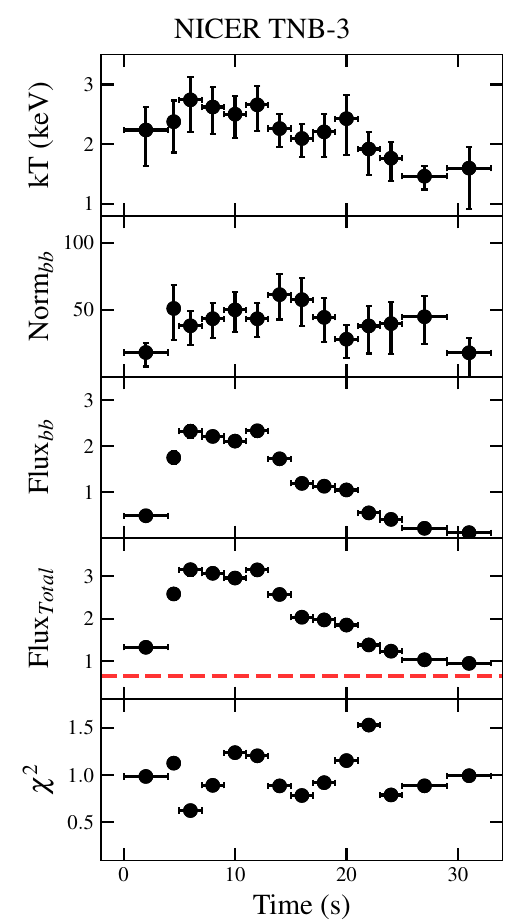} 
       \includegraphics[width=0.65\columnwidth]{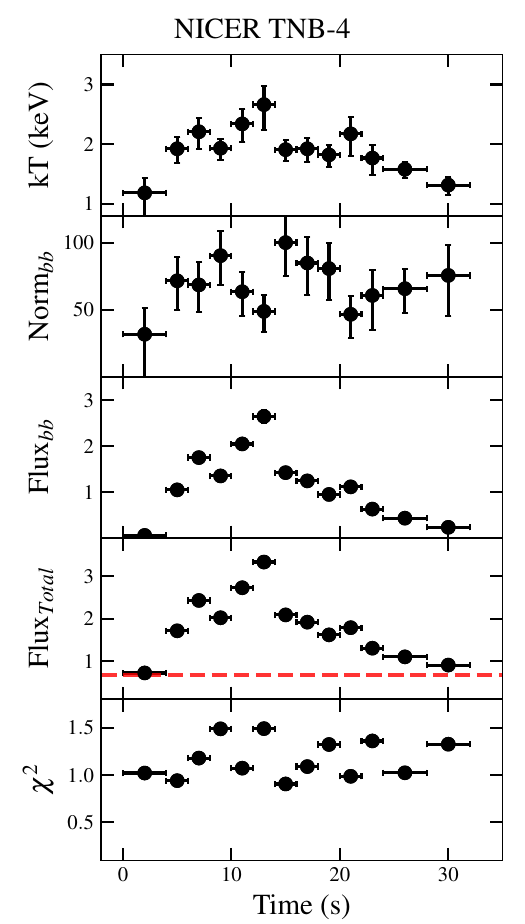}        
\caption{The evolution of different spectral parameters during bursts obtained by fitting time-resolved \nicer{} spectra of \obj{} with model {\tt tbabs$\times$bbodyrad}. The blackbody temperature (kT; top panel), blackbody normalization (second panel), blackbody flux (third panel), total flux (fourth panel), and reduced $\chi^2$ (bottom panel) over the progress of the bursts are shown. The X-axis indicates the time from the rise of the bursts. The fluxes are absorption corrected and estimated in the 0.1--100 keV ranges and quoted in the unit of 10$^{-8}$ erg cm$^{-2}$ s$^{-1}$. The red horizontal dashed line indicates the persistent flux level.}
    \label{fig:fig_time_Resolved_nicer}
\end{figure*}

\section{Results}
\label{res}
 We carried out a detailed temporal and spectral study of the AMSP \obj{} using \nicer, \xmm{}, and \nustar{} observations. The source was monitored with various X-ray observatories during the 2024 outburst. Figure~\ref{fig:MAXI} shows the daily monitoring light curve of the source with {\it MAXI}/GSC. The arrows in the figure indicate the epochs of pointed X-ray observations of the source using different instruments in this study. In the \nicer{} observations, five thermonuclear bursts are detected in the light curves. Among the five bursts, one is not covered entirely, and another burst (\nicer{} TNB-1) has already been reported by \citet{Ng24}. 
 In this work, we performed a detailed analysis of the newly found 3 \nicer{} bursts, and \nicer{} TNB-1 is also used to investigate the reflection signature in the burst. During \xmm{} and \nustar{} observations, many bursts were detected, among which nine bursts were observed simultaneously. Although one burst was detected in the \swift{} observation, due to the poor signal-to-noise ratio for time-resolved spectral study, a detailed spectral analysis was not performed. We generated burst profiles across different energy ranges and conducted time-resolved spectral studies during the bursts. In addition, we also investigated the spectral features during the burst-free persistent emission regions using the \nicer{}, \xmm{}, and \nustar{} observations. 


\subsection{Burst profile and its energy dependence}

Thermonuclear bursts from \obj{} were detected with different instruments during the 2024 outburst. Figure~\ref{fig:burst_combined} shows the light curves of \obj{} using different instruments. It can be seen that several bursts were detected with \nicer{}, \xmm{}, and \nustar{}, which provides a unique opportunity to investigate the different properties of the source both in broadband during the bursts and in burst-free, persistent emission regions. The details of the bursts are summarized in Table~\ref{tab:log_table_burst}. The thermonuclear bursts observed with \nicer~ \& \xmm~ are for durations of $\sim$25--30~s with burst rise and decay times of $\sim$3-5~s and $\sim$10-15~s, respectively, along with a plateau of $\sim$6~s. However, the bursts observed with \nustar~ are of a duration of $\sim$20-25~s, with rise and decay times of $\sim$5~s and $\sim$10-15~s, respectively.

To investigate the energy dependence of the burst profiles, we extracted light curves in different energy ranges such as 0.5-3 keV, 3-5 keV, 5-10 keV, 0.5-5 keV, and 0.5-10 keV for \nicer{} and \xmm{} EPIC-PN. For the \nustar{} observation, energy-resolved light curves are extracted for energy ranges of 3-5 keV, 5-10 keV, 10-20 keV, and 3-79 keV. The energy dependence of the burst profiles is investigated for each burst observed with \nicer{}, \xmm{}, and \nustar{}. A strong energy dependence of the burst profiles is observed for all bursts. The bursts are detected up to $\sim$20 keV with \nustar{}. Although the rise time remains comparable across all energy bands, the bursts decay faster at higher energies and exhibit a relatively longer tail at lower energies. Figure~\ref{fig:burst_profile} shows the energy-resolved profiles for different bursts from \nicer{}, \xmm{}, and \nustar{} observations. The peak count rate for the \nicer{} (0.5-10 keV) and \xmm{} EPIC-PN (0.5-10 keV) bursts are $\sim$1000 counts s$^{-1}$ and $\sim$1200 counts s$^{-1}$, respectively. During \nustar{} (3-79 keV) bursts, the peak count rate is nearly 800 counts s$^{-1}$. For the energy-resolved profiles of all bursts, see Figure~\ref{fig:burst_profile_all}. 


\begin{table*}
\centering
 \caption{Best-fit spectral parameters [{\tt XSPEC} model : $\texttt{TBabs}\times( \texttt{po}+\texttt{diskbb})$] of the pre-burst \nicer{} (0.5--10 keV) and {\xmm{} EPIC-PN + \nustar{}} (0.5--25 keV) spectra of \src{}.}
\begin{tabular}{llccccc}
\hline
Components & Parameters & \nicer{} & \nicer{} & \nicer{} & \nicer{} & \xmm{}+\nustar{}  \\
 &  & TNB1 & TNB2 & TNB3 & TNB4 & TNB   \\
\hline
{\tt tbabs} & N$_{\textrm{H}}$ & $2.8\pm0.3$ & 2.8$\pm0.1$ & 2.8$\pm0.2$ & 2.9$\pm0.2$ & $2.6\pm0.2$  \\
\hline
{\tt power-law} & $\Gamma$ & $1.6\pm0.5$ &1.7$\pm$0.2 & 1.8$\pm$0.2 & 1.8$\pm$0.2 & 1.7$\pm0.1$  \\
  & Norm. & $0.4\pm0.3$ &0.4$\pm$0.1 & 0.4$\pm$0.2 & 0.4$\pm$0.2 & 0.3$\pm0.1$  \\
\hline
{\tt diskbb} & T$_{\textrm{in}}$(keV) & $1.2\pm0.5$& 1.1$\pm$0.2 & 0.9$\pm$0.1 & 0.9$\pm$0.2 & 1.6$\pm$0.2  \\
  & Norm. & 23$\pm$25 &20$\pm$13 & 43$\pm$30 & 29$\pm$25 & 5$\pm$3 \\
\hline
 & Flux$_\mathrm{Total}$  & 10.0$\pm$1.0 & 8.6$\pm 0.1$ & 6.4$\pm 0.5$ & 6.8$\pm$ 0.2 & 5.0$\pm$0.2  \\
  & Luminosity  & $11.9 \pm1.2$ & 6.6$\pm$ 0.1 & 4.9$\pm$ 0.4 & 5.2$\pm 0.2$ & 3.8$\pm 0.2$ \\
 & $\dot{m}$  & $6.7\pm0.5$ & 5.8$\pm$0.1 & 4.3$\pm$0.3 & 4.6$\pm$0.1 & 3.4$\pm$0.1   \\
 & $\dot{m} / \dot{m}_{\mathrm{Edd}}$  & $\sim$0.76& $\sim$0.65 & $\sim$0.49  & $\sim$0.52 & $\sim$0.39  \\
\hline
 & $\chi^{2}$/dof & 291/300 & 660/610 & 564/506 & 525/510 & 665/616  \\
\hline
\label{tab:tab_preburst}
\end{tabular}\\
 The reported errors are of 90\% significant. All fluxes (0.1--100 keV range) are corrected for absorption and quoted in the unit of 10$^{-9}$ ergs cm$^{-2}$ s$^{-1}$. The X-ray luminosity is in the units of 10$^{37}$ erg s$^{-1}$ (assuming a source distance of 10 kpc), $N_\textrm{H}$ is in the unit of $\rm 10^{22}~ cm^{-2}$. The mass accretion rates $\dot{m}$, $\dot{m}_{\mathrm{Edd}}$ are in the unit of 10$^{4}$~g~cm$^{-2}$ s$^{-1}$, The Eddington rate ($\dot{m}_{\mathrm{Edd}}$) for a typical NS of radius 10 km is assumed to be $8.8\times10^{4}$ g cm$^{-2}$ s$^{-1}$ \citep{Ga08}. 
\end{table*}


\begin{figure*}
  \includegraphics[width=0.5\columnwidth]{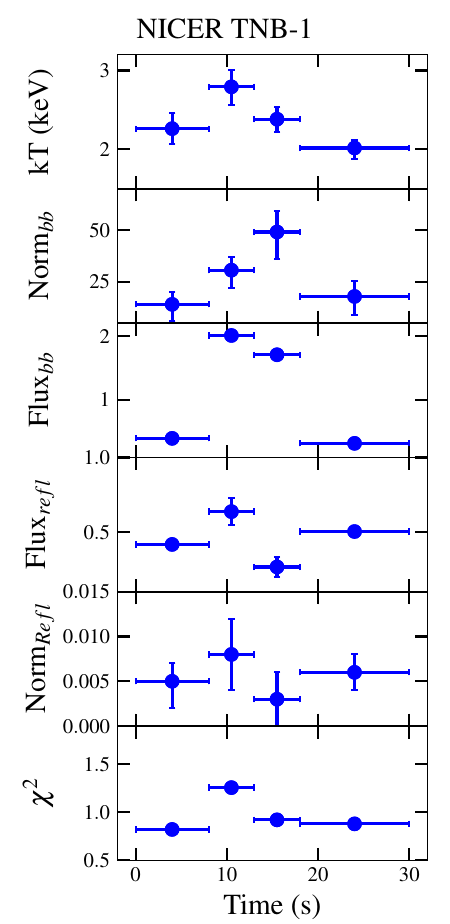}
  \includegraphics[width=0.5\columnwidth]{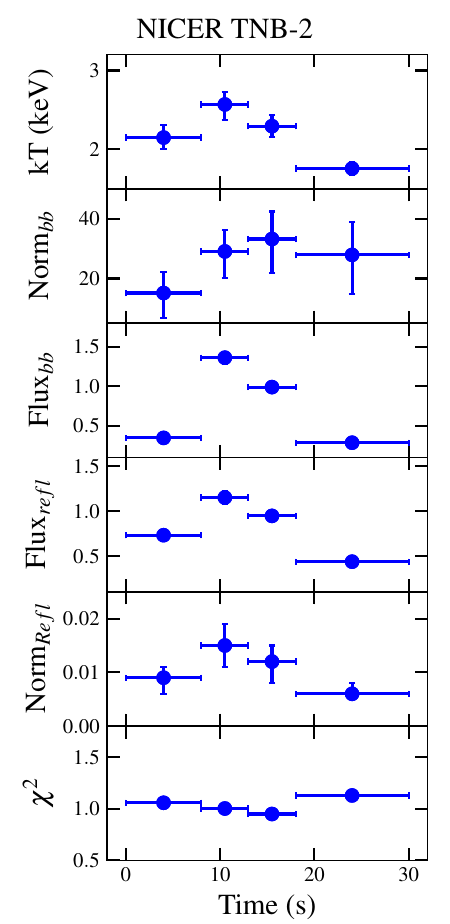}
  \includegraphics[width=0.5\columnwidth]{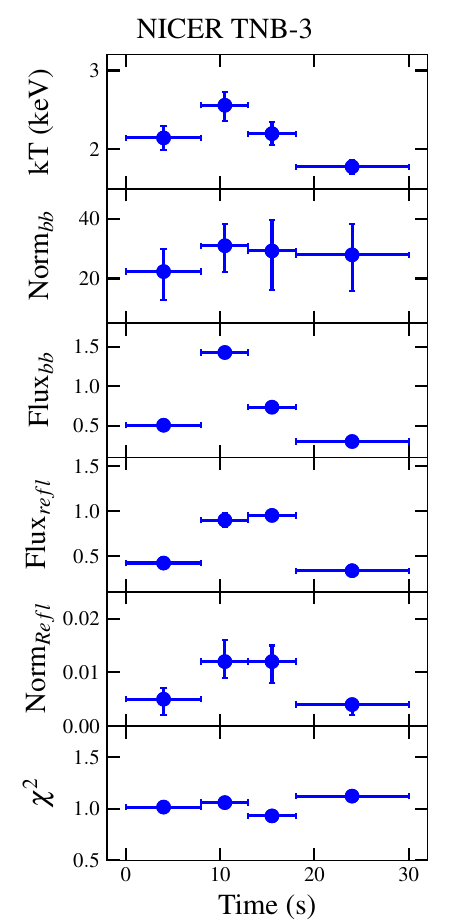}
    \includegraphics[width=0.5\columnwidth]{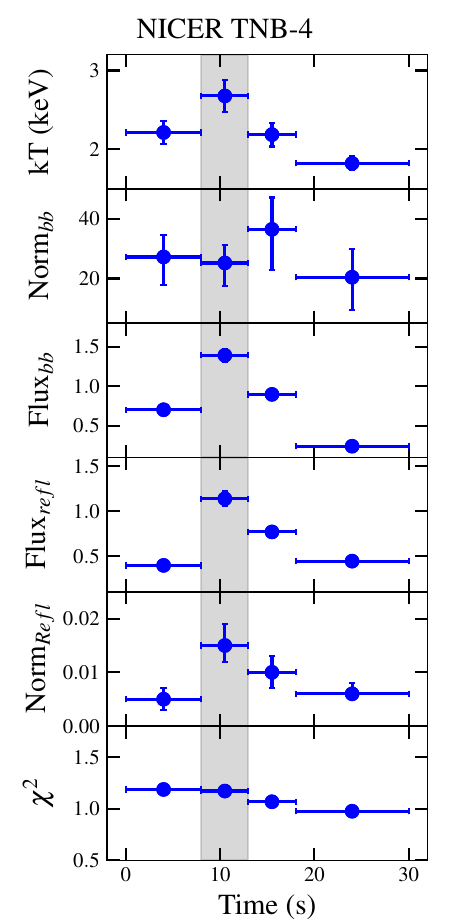}
 
\caption{Evolution of different spectral parameters during bursts obtained by fitting time-resolved \nicer{} spectra of \obj{} with model {\tt tbabs $\times$ (bbodyrad + relxillNS)}: blackbody temperature (kT; Top panel), blackbody normalization (second panel), blackbody component flux (third panel), reflection component flux (fourth panel), normalization of {\tt relxillNS} (fifth panel), and reduced $\chi^{2}$ (bottom panel). Measured fluxes are in the energy range of 0.1--100 keV and quoted in the unit of 10$^{-8}$ erg cm$^{-2}$ s$^{-1}$. The X-axis indicates the time from the rise of the burst. The shaded region indicates the segments for burst peak, and the corresponding spectrum is shown in Fig. \ref{fig:reflection_nicer}.}
    \label{fig:nicer_reflection}
\end{figure*}


\begin{figure}
   \includegraphics[width=\columnwidth]{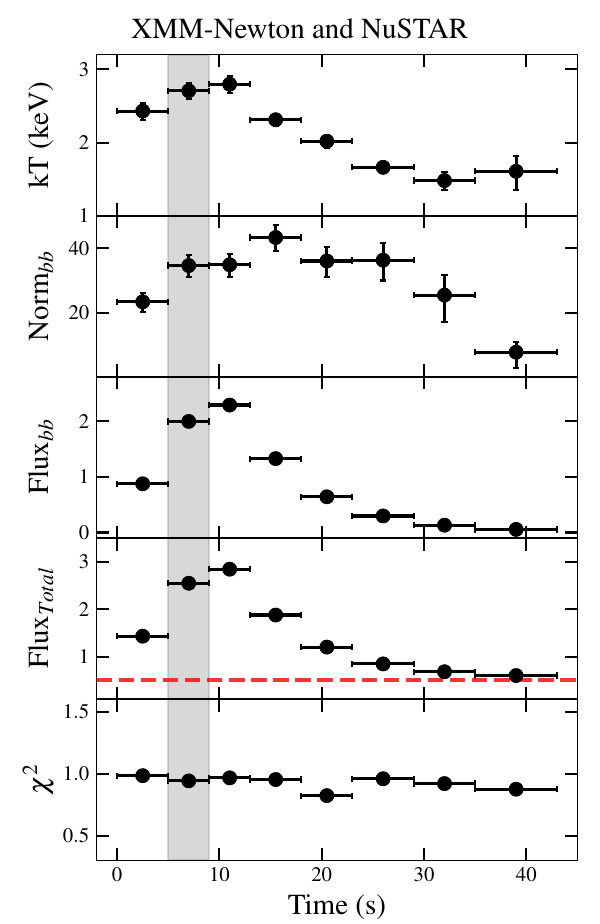} 
\caption{Evolution of various spectral parameters obtained from the time-resolved simultaneous spectral fitting of a burst from \obj{} observed with  \xmm{} and \nustar{} (marked with asterisks in third and fourth panels of Figure~\ref{fig:burst_combined}) with the model {\tt tbabs $\times$ bbodyrad}. The variation of blackbody temperature (kT; first panel), blackbody normalization (second panel), blackbody flux (third panel), total flux (fourth panel), and reduced $\chi^2$ (bottom panel) are shown. The X-axis presents the time from the rise of the TNB. The fluxes (in 0.1--100 keV range) are corrected for absorption and quoted in the unit of 10$^{-8}$ erg cm$^{-2}$ s$^{-1}$. The red horizontal dashed line indicates the persistent flux level.}
    \label{fig:fig_time_Resolved_nustar_xmm}
\end{figure}

\begin{figure*}
    \includegraphics[width=0.725\columnwidth, angle=270]{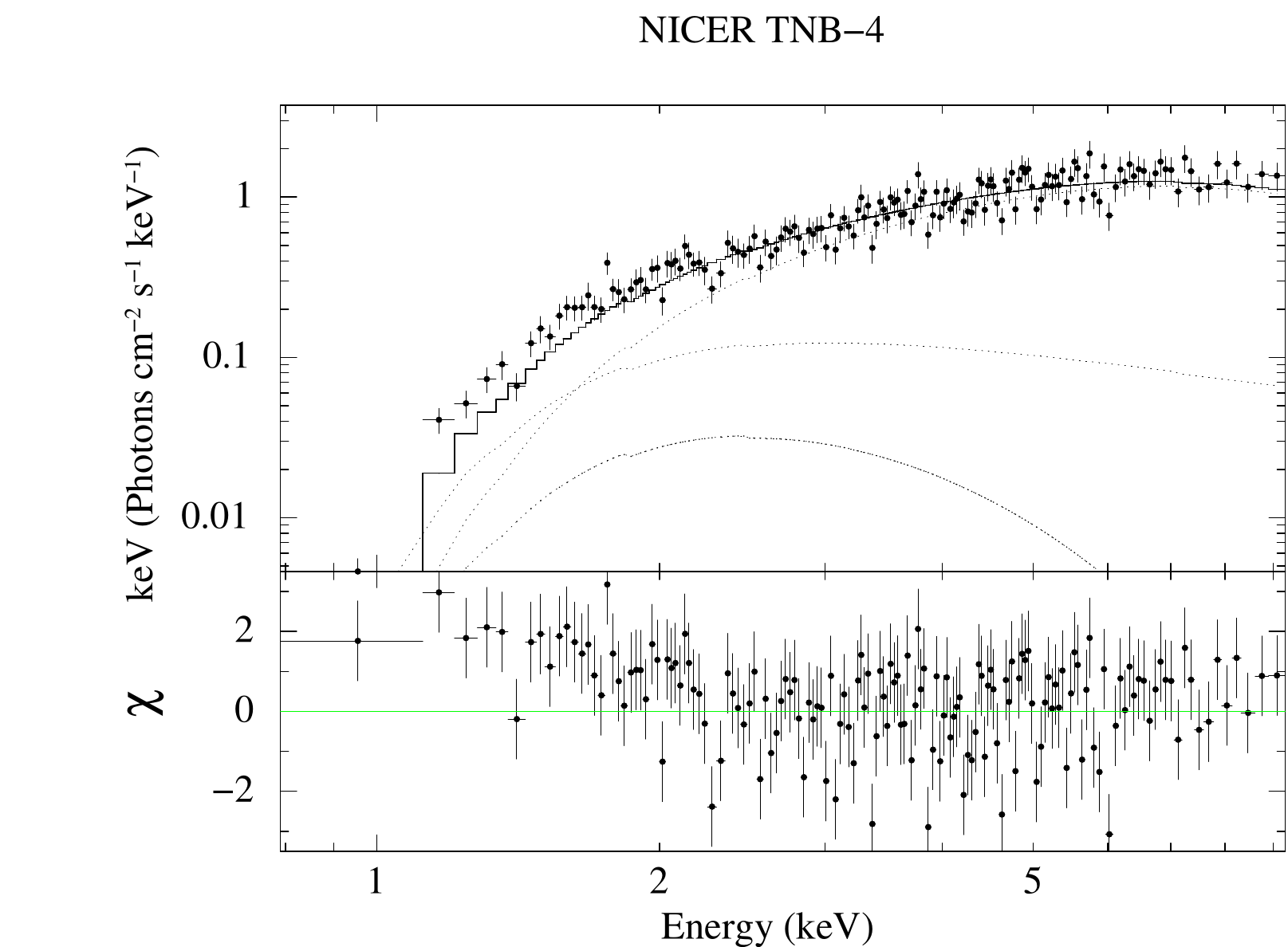} 
    \includegraphics[width=0.725\columnwidth, angle=270]{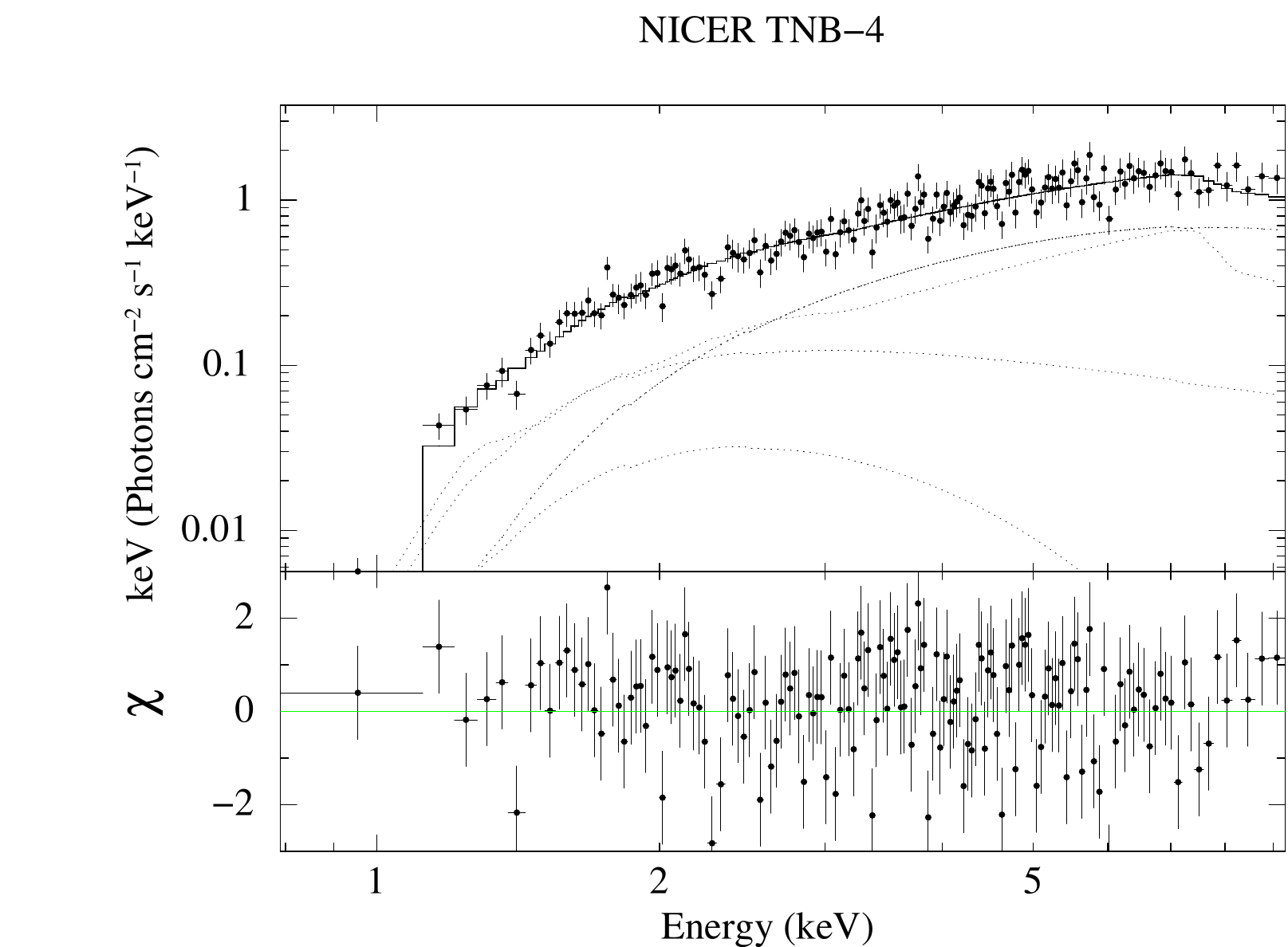} 
\caption{{\it Left}: The spectrum of \obj{} at the peak of the \nicer{} burst (TNB-4; shaded strip in the right panel of Figure~\ref{fig:nicer_reflection}) fitted with the absorbed blackbody model, showing the presence of a soft excess below 2 keV. {\it Right}: The spectrum at the peak of the \nicer{} burst (TNB-4) fitted with the absorbed blackbody and an additional reflection model \texttt{relxillNS} is shown in the top panel. Corresponding residuals are shown in the bottom panel. The addition of the reflection component is found to be statistically significant (F-test: F value = 37.8 and chance probability of 10$^{- 9}$.}
    \label{fig:reflection_nicer}
\end{figure*}

\begin{figure*}
    \includegraphics[width=0.725\columnwidth, angle=270]{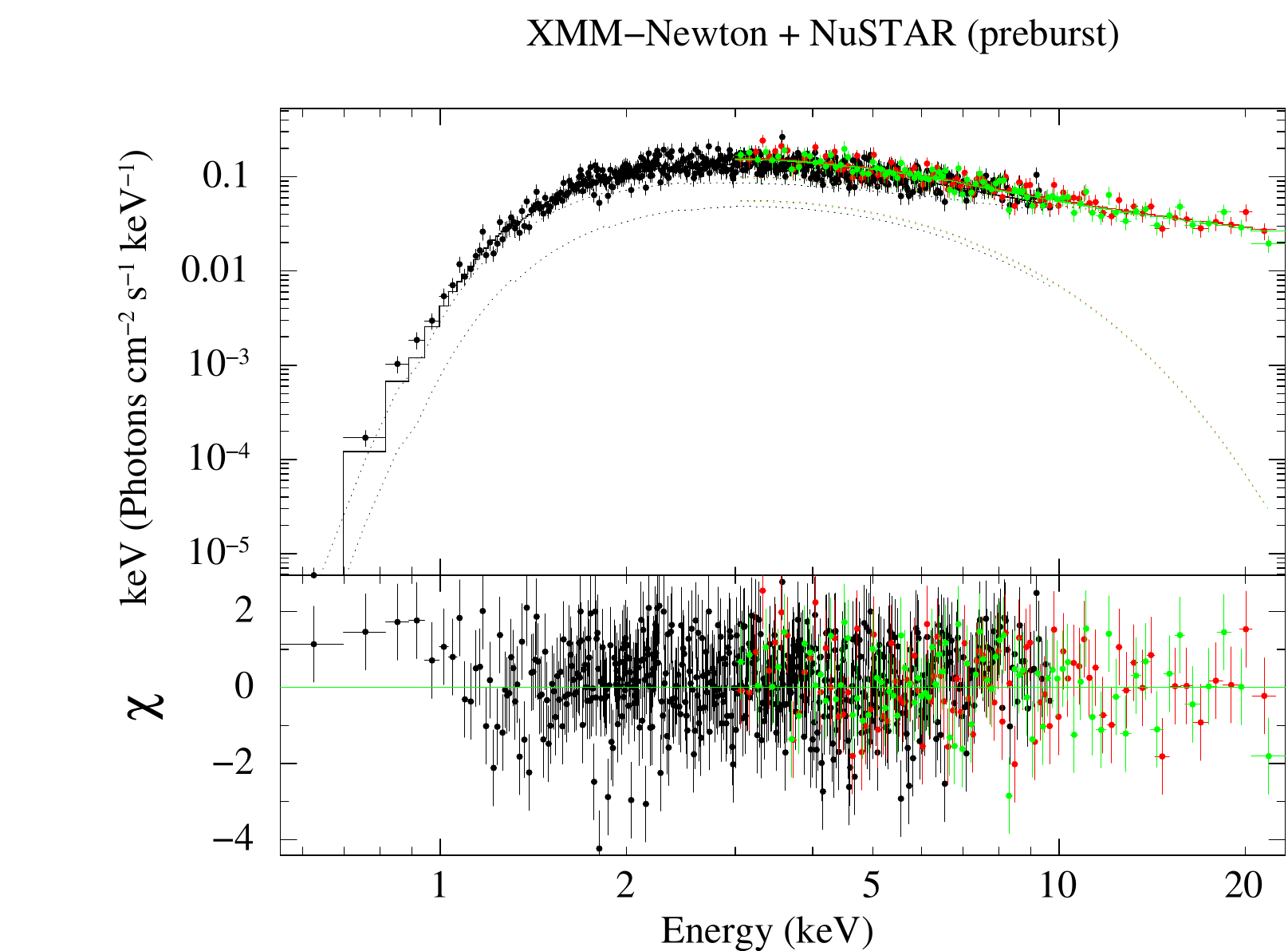} 
    \includegraphics[width=0.725\columnwidth, angle=270]{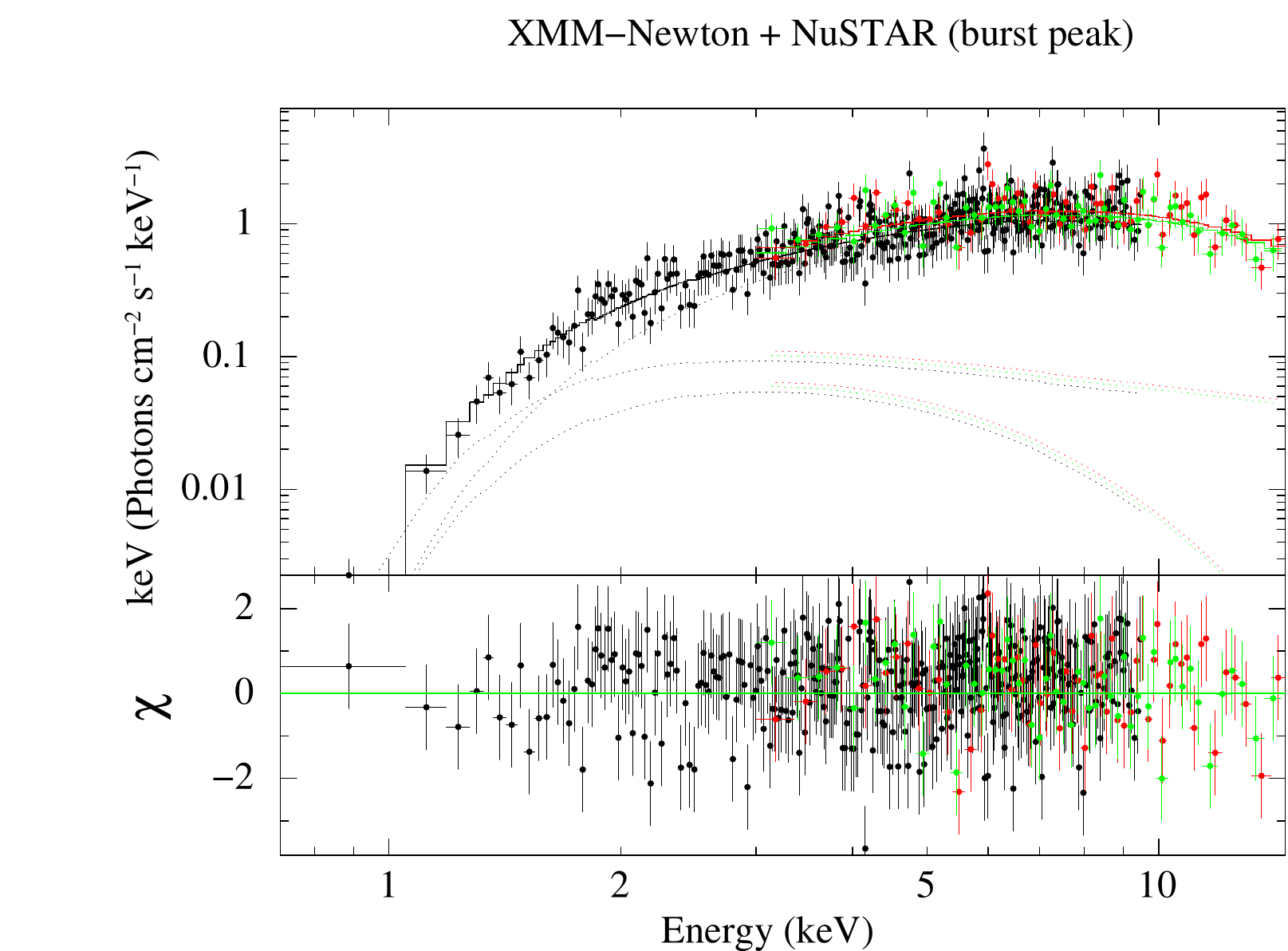} 
\caption{{\it Left}: The broadband \xmm{} and \nustar{} pre-burst spectra fitted with the absorbed power law and a disk blackbody model. {\it Right}: The best fit broadband spectra with the absorbed blackbody model during the burst peak (corresponding to the shaded segment of Figure~\ref{fig:fig_time_Resolved_nustar_xmm}), and corresponding residuals are shown in the bottom panels, respectively.}
    \label{fig:spec_XMM_NuSTAR}
\end{figure*}

\subsection{Time-resolved burst spectral analysis}
\label{time_resolved_spectroscopy}
Time-resolved spectroscopy is performed to probe the dynamic evolution of the X-ray bursts observed with \nicer{}. The pre-burst \nicer{} spectra are obtained for a duration of 200 s. We also carried out time-resolved spectroscopy for a burst simultaneously observed with \xmm{} and \nustar{} (marked with asterisks in Figure~\ref{fig:burst_combined}) to probe the broadband emission mechanism during the burst. As in the case of \nicer{}, pre-burst spectra for \xmm{} and \nustar{} bursts are also extracted for a duration of 200 s. The pre-burst spectra are modeled with a power law continuum and a multi-temperature disk blackbody ({\tt diskbb}) model in {\tt XSPEC} \citep{Ar96}. The interstellar absorption is accounted for using the {\tt TBABS} \citep{Wilms2000} model. The best-fit pre-burst spectral fitting parameters are presented in Table~\ref{tab:tab_preburst}. The pre-burst fitting results provide a photon index of 1.7, an inner disk temperature of 0.9-1.6 keV, and the hydrogen column density ($N_\textrm{H}$) of $\sim2.8 \times 10^{22}$ cm$^{-2}$.

A total of 13 spectra are created for each burst observed with \nicer{} to conduct the time-resolved spectral study. The \nicer{} spectra are generated in the time bin of 2 s, whereas the initial and last two segments during the burst are created for 4-s segments. In the burst decay phase, the time bin is extended from 2\,s to 4\,s to maximize the signal-to-noise ratio. We also performed a time-resolved spectral study for a burst detected simultaneously with \xmm{} and \nustar{}. The burst is divided into eight segments, and the corresponding spectra and responses are generated for both the \xmm{} and \nustar{} instruments. The burst rising part (1st segment) is for 5\,s duration, the 2nd and 3rd segments are for the 4\,s durations, the 4th and 5th segments are for 5~s durations, and segments 6, 7, and 8 are for 6~s, 6~s, and 8~s, respectively. A comparatively longer time bin is selected in the rising and decay phases to optimize the signal-to-noise ratio based on the total counts available for a reasonable time-resolved broadband spectroscopy.

The time-resolved burst spectra are modeled with an absorbed blackbody model ({\tt bbodyrad}). As low-energy photons are affected by absorption, and the burst segments have shorter durations, we used data in the range of 1-10 keV for our time-resolved spectral modeling. During the fitting, to consider the persistent emission, we use the values of the best-fit spectral parameters for pre-burst segments (see Table~\ref{tab:tab_preburst}). We do not find any requirement for an additional scaling factor to fit the burst spectra. The evolution of spectral parameters during the bursts detected in \nicer{} observations is shown in Figure~\ref{fig:fig_time_Resolved_nicer}. The blackbody temperature reaches a maximum value of $\sim$3 keV with the 0.1-100 keV peak blackbody flux in the range of (2.3-2.7) $\times 10^{-8}$ erg cm$^{-2}$ s$^{-1}$. We do not find any signature of PRE during any of the bursts observed with \nicer{}. The peak flux during the bursts is mentioned in Table~\ref{tab:log_table_burst}. Typically, a soft excess emission is observed in the burst spectrum. However, for this source, excess emission in soft X-rays is not evident, likely due to the absorption of low-energy photons (below 1 keV) by a high column density \citep{Ng24, Gu21}. To increase the low energy count statistic ($\leq$1 keV), we also used comparatively longer segments for time-resolved spectral study during the burst (Section~\ref{reflection_burst}).

\begin{figure}
\centering{
\includegraphics[width=8cm]{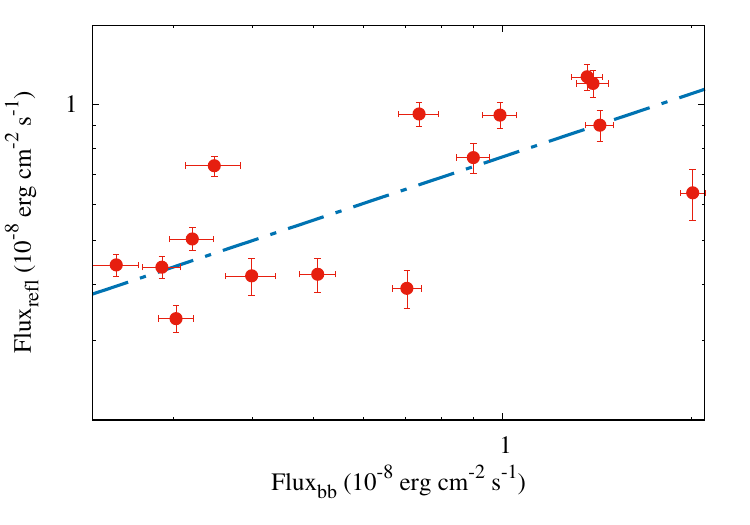}
\caption{Correlation between the blackbody and reflection component flux of \src{} during the \nicer{} bursts. 
}
\label{fig:correlation}}
\end{figure}

We investigate the dynamic evolution of spectral parameters in a broad energy range by analyzing a simultaneously detected thermonuclear burst using \xmm{} and \nustar{} data. The burst is divided into 8 segments to optimize the signal-to-noise ratio. To conduct the time-resolved spectral analysis of the simultaneously detected burst with \nustar{} and \xmm{}, we created spectra and responses for segments of the burst by generating GTIs for each segment of both data sets. The broadband time-resolved burst spectra from \xmm{} EPIC-PN, and \nustar{} FPMA/FPMB are fitted simultaneously for each segment. We use the same model for the broadband \xmm{} and \nustar{} time-resolved spectral modeling as in the case of \nicer{} bursts. In addition, a multiplicative constant is used to account for the uncertainties of cross-calibration of different instruments. The multiplicative constant is fixed at 1 for \xmm{} EPIC-PN and allowed to vary freely for FPMA and FPMB of \nustar{}. The multiplicative constant is found to be in the range of 1.03-1.2 and 0.9-1.1 for FPMA and FPMB, respectively, for different segments. Figure~\ref{fig:fig_time_Resolved_nustar_xmm} shows the evolution of spectral parameters during a thermonuclear burst simultaneously detected with the \xmm{} EPIC-PN and \nustar{}. Figure~\ref{fig:spec_XMM_NuSTAR} shows the broadband best-fit pre-burst spectra (left panel) and the burst peak spectra (right panel) from the \xmm{} and \nustar{} observations. During the peak of the burst, the blackbody temperature and the blackbody flux (0.1--100 keV range) reached maximum values of 2.8 keV and $\sim2.3\times10^{-8}$ erg cm$^{-2}$ s$^{-1}$, respectively.

\subsubsection{Disk reflection during thermonuclear bursts}
\label{reflection_burst}

During thermonuclear bursts, a fraction of the burst photons may interact with the disk and be reflected by the disk, depending on the disk inclination and inner disk radius \citep{Ga21, Ke18a}. To investigate this feature, we generated burst spectra in 4 segments of a comparatively longer time duration to optimize the signal-to-noise ratio for each \nicer{} burst. We used the disk reflection model to investigate further the burst spectra, which show a deviation from a pure blackbody model (see Figure~\ref{fig:reflection_nicer}). During the bursts, the reflection feature is probed using the latest relativistic reflection model {\tt relxillNS}. This model assumes that the photoionized disk is illuminated by a blackbody spectrum from the NS \citep{Garcia2022}. The {\tt relxillNS} model consists of following model parameters: inner disk radius ($R_{\textrm{in}}$), outer disk radius ($R_{\textrm{out}}$), ionization parameter ($\log{\xi}$), disk density ($\log{N}$), iron abundance ($A_{\textrm{Fe}}$), dimensionless spin parameter ($a^\ast$), emissivity indices (q1, q2), and the input blackbody temperature ($kT_{\textrm{bb}}$). As the reflection model consists of many parameters, it is challenging to constrain all parameters while fitting data from small-duration segments. The disk parameter values in the {\tt relxillNS} model are chosen based on prior studies of similar systems \citep{Lu19, Zh22, Lu23, Yu24, 4U1702, Ma25}. The model parameters are set as following: emissivity index q1 = q2 = 3, $R_{\textrm{in}}$ = $\rm R_{\textrm{ISCO}}$, $R_{\textrm{out}}$ = 400$R_g$, $i$ = 50$^{\circ}$, $A_{\textrm{Fe}}$ = 5, $\log{\xi}$ = 3.2 erg~cm~s$^{-1}$, $\log{N}$ = 18 cm$^{-3}$. The input blackbody temperature is tied with the temperature of the {\tt bbodyrad} model. The \nicer{} time-resolved burst spectra can be fitted using the model {\tt tbabs $\times$~(bbodyrad + relxillNS + powerlaw + diskbb)}, where ({\tt powerlaw + diskbb}), {\tt relxillNS}, {\tt bbodyrad} represent the contribution of persistent emission, reflection of burst photons, and the burst, respectively. During the modeling of time-resolved burst spectra, we used the pre-burst model parameter values as obtained from pre-burst spectral modeling (see Table~\ref{tab:tab_preburst}). The reflection fraction is set to a negative value of -1, so that the {\tt relxillNS} describes the reflection component and the {\tt bbodyrad} describes the direct coronal emission. The left panel of Figure~\ref{fig:reflection_nicer} shows an excess in soft energies when the burst peak spectrum (shaded strip in the right panel of Figure~\ref{fig:nicer_reflection}) is fitted with an absorbed blackbody model. The disk reflection model is used to describe this feature, and the best-fit spectrum is shown in the right panel of Figure~\ref{fig:reflection_nicer}. To check the statistical significance of the reflection component, the F-test is performed. The F-test provides the $F$ value and a chance probability of 37.8 and $10^{-9}$, respectively. The reflection model indicates that the probability of a decrease in $\chi^2$ value is less than 5\%. Therefore, the reflection model is used to investigate the evolution of burst spectral parameters along with the normalization and flux of the reflection component. Figure~\ref{fig:nicer_reflection} shows the evolution of different spectral parameters during the bursts observed with \nicer{}. The peak blackbody temperature is found to be $\sim$2.8 keV, and the peak blackbody flux (0.1--100 keV range) is found to be in the range of 0.3--2.2~$\times$~10$^{-8}$ erg~cm$^{-2}$~s$^{-1}$ for \nicer{} bursts. The flux of the reflection component is in the range of 0.3--1.2~$\times$~10$^{-8}$ erg~cm$^{-2}$~s$^{-1}$ at the peak of \nicer{} bursts. It is found that the reflection component is contributing up to a maximum of 30\% of the total flux as observed for the \nicer{} bursts. We also investigated the correlation between the flux from the blackbody and reflection components during these \nicer{} bursts and showed in Figure~\ref{fig:correlation}. A positive correlation is observed between the reflection and blackbody component fluxes. The best-fit result with a function $F_{\textrm{refl}} = k~F_{\textrm{bb}}^{\alpha}$ provides k = 0.7 $\pm$ 0.1 and $\alpha$ = 0.5 $\pm 0.1$. The pre-burst flux during \xmm{} observation was lower ($\sim$40\%) compared to the \nicer{} observations. The peak flux during the \xmm{} burst was also lower than that of the \nicer{} bursts. This may be the possible reason that the reflection feature was not detectable during the \xmm{} burst.


\subsection{Modeling of persistent spectra: Reflection}
\label{reflection_persistent}

\begin{figure}
\includegraphics[width=0.65\columnwidth, angle=270]{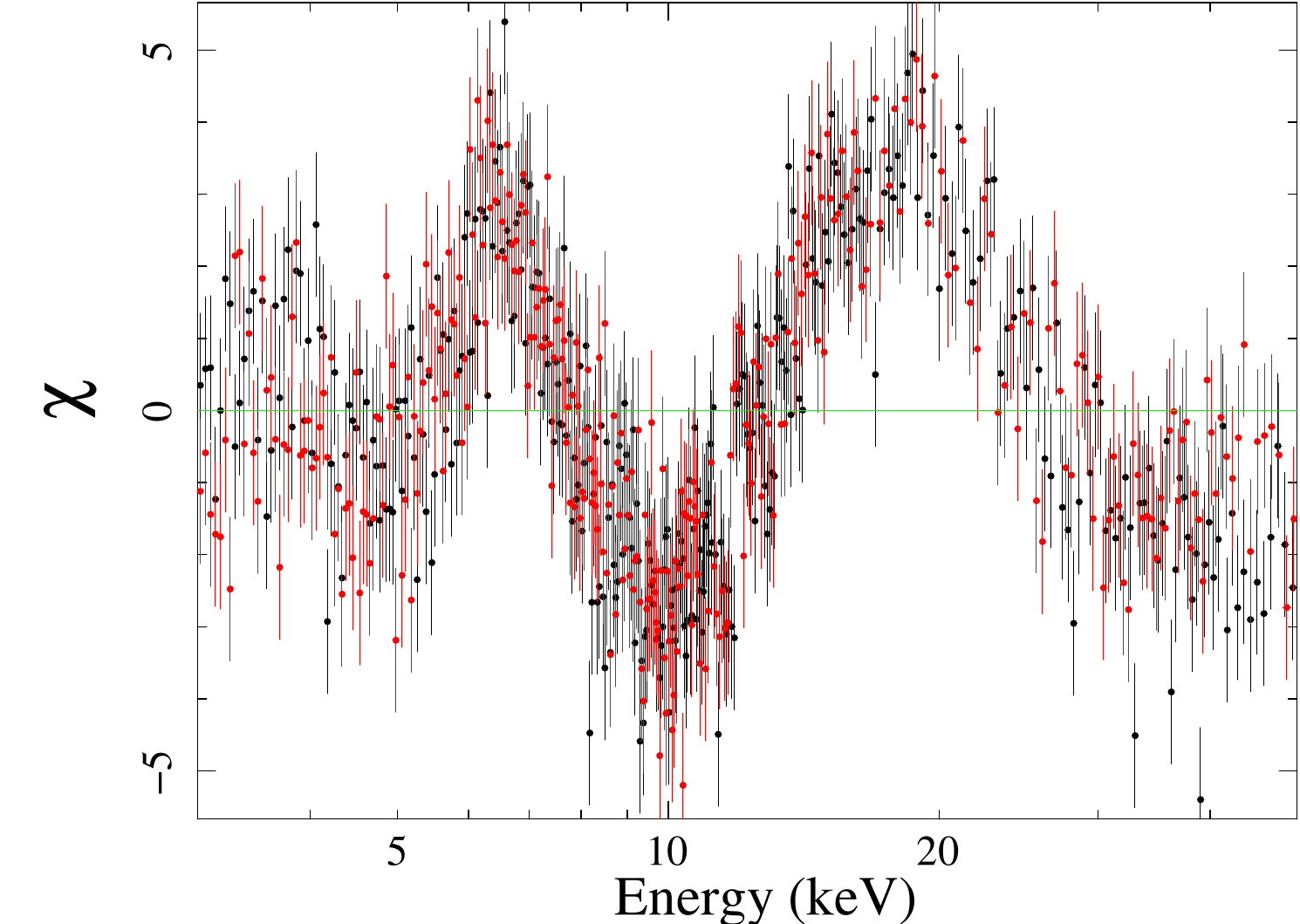}

\caption{The residuals of the burst-free \nustar{} spectra of \obj{} fitted with an absorbed cutoff power law and blackbody model {\tt tbabs $\times$ (cutoffpl + bbodyrad)}. The presence of an iron line at $\sim$6.4 keV and a Compton hump near 20 keV is evident in the residuals.}
    \label{fig:delchi}
\end{figure}

 We investigated reflection features by considering simultaneous burst-free broadband data from the \nicer, and \nustar{} observations of \obj{}. The spectra and response matrices are generated from \nicer~ and \nustar~ observations by creating GTI files for the region between the vertical dotted lines, as shown in Figure~\ref{fig:burst_combined}, for simultaneous spectral fitting. To model the broadband energy spectra methodically, we first applied single continuum models, such as the power-law model modified by the {\tt TBabs} model, the single temperature blackbody component ({\tt bbodyrad} in {\tt XSPEC}), the thermal Comptonization component ({\tt nthComp} in XSPEC: \citep{Zd96, Zy99}), and the multi-temperature disk blackbody component ({\tt diskbb} in {\tt XSPEC}: \citet{Mi84, Ma86}). The line of sight absorption due to Galactic neutral hydrogen is accounted for using the {\tt TBabs} model \citep{Wilms2000} provided in the {\tt XSPEC}. The most recent abundance model, {\tt wilm}, is used and incorporated into {\tt XSPEC} for the initial input abundance in the {\tt TBabs} model. We include a constant multiplier factor to account for uncertainties related to the cross-instrument calibration. The constant was fixed at a value of 1 for \nustar{}/FPMA, while for \nustar{}/FPMB, and \nicer{}, it was allowed to vary. A higher reduced $\chi^2$ value is obtained while modeling using these single continuum models. To describe the broadband spectra, we proceeded to introduce relatively sophisticated three-component models, such as {\tt TBabs $\times$ (bbodyrad + cutoffpl)} and {\tt TBabs $\times$ (bbodyrad + nthComp)}. The fitting statistic under the {\tt TBabs $\times$ (bbodyrad + cutoffpl)} model is improved compared to the previous model combination. We found that the residuals (see Figure~\ref{fig:delchi}) contain X-ray reflection features, including emission features in the energy range of 5-8 keV, a back-scattering Compton hump at $\sim$20 keV, and a photoelectric absorption dip at $\sim$10 keV, regardless of the continuum component selection.


\begin{figure}
\includegraphics[width=0.72\columnwidth, angle=270]{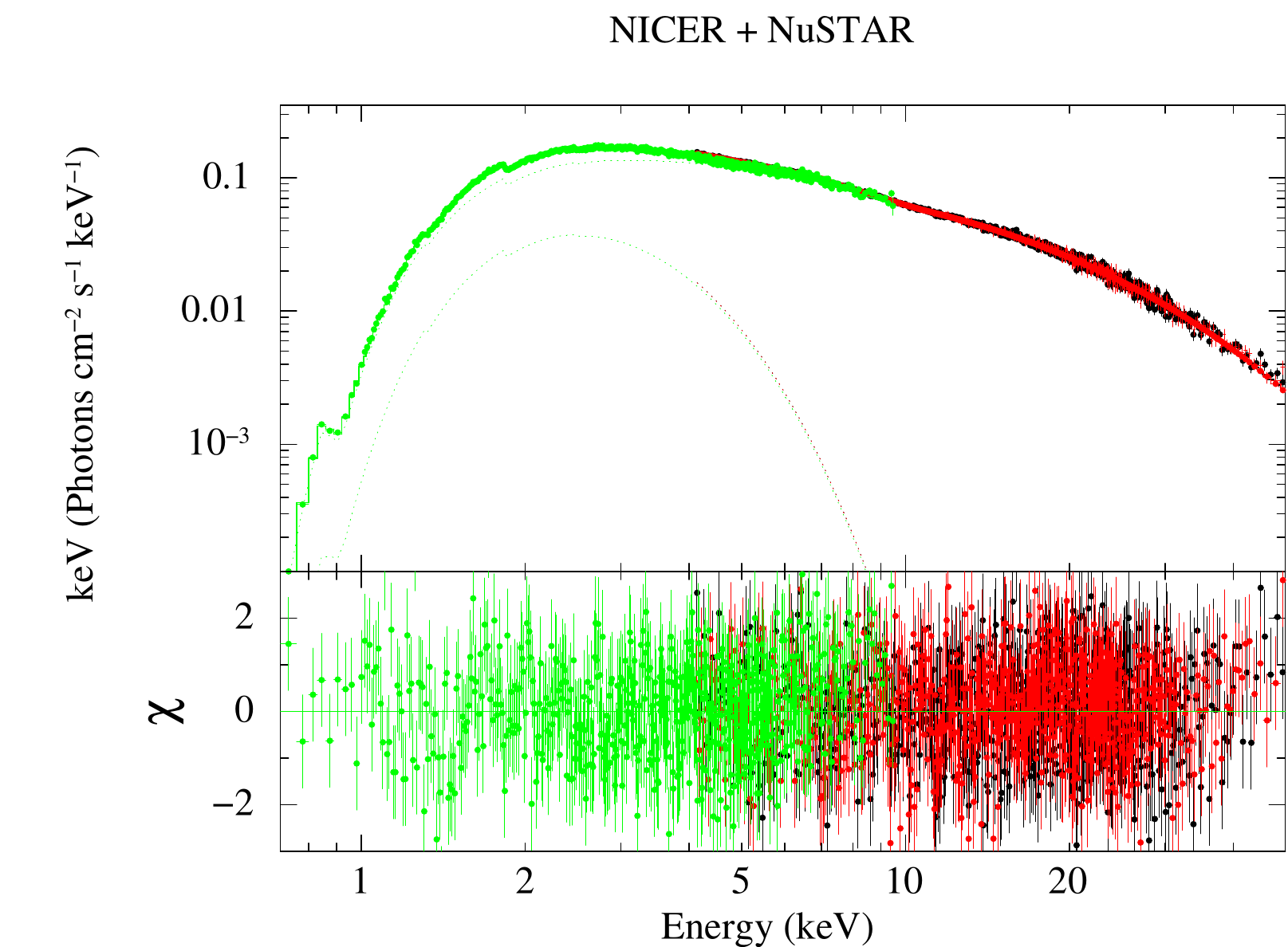}
\caption{The broadband burst-free \nicer{} (in green color) and \nustar{} spectra of \obj{} with the best-fit model \texttt{tbabs $\times$ (relxill + bbodyrad)}, and corresponding residuals are shown in bottom panel. The \nustar{} FPMA and FPMB spectra are shown in black and red colors, respectively.}
    \label{fig:reflection_nustar_nicer}
\end{figure}

 We used the {\tt relxill} model, a flavor of the broader {\tt relxill} model family (v2.3; \citet{Ga14, Da14}), which models relativistic reflection of a cut-off power-law continuum from the accretion disk. In the {\tt relxill} scenario, the primary source is assumed to have emissivity Index1 (q1) and Index2 (q2). The reflection model also consists of parameters such as the photon index ($\Gamma$) of the illuminating radiation, cutoff energy ($E_{\textrm{cut}}$), disk inclination angle ($i$), inner disk radius ($R_{\textrm{in}}$), outer disk radius($R_{\textrm{out}}$), dimensionless spin parameter($a$), the ionization parameter ($\xi$) at the surface of the disk, and iron abundance relative to its solar value ($A_{\textrm{Fe}}$). The broadband spectrum (0.7-50 keV) is fitted with the model {\tt TBabs $\times$ (relxill + bbodyrad)}, which significantly improves the fitting results (reduced $\chi^2 \sim$1). The {\tt relxill} model includes the cutoff power-law model {\tt cutoffpl} as an illuminating continuum. For modeling of the broadband spectra, the emissivity indices are set as q1 = q2 = 3, $R_{\text{out}}$ is fixed at 400 gravitational radii ($R_g$ = GM/$c^2$). \citep{Ng24} detected a pulsation of 447.9 Hz, which is used to calculate the dimensionless spin parameter a$^\ast$ using the relation a$^\ast$ $\simeq$ 0.47/P (ms) \citep{Braje2000}. The dimensionless spin parameter ($a^\ast$) is frozen at 0.21 during the spectral modeling. Two edges at 0.9 keV and 1.8 keV are included in the spectral model for the known features in the instrument calibration in \nicer{}\footnote{\url{https://heasarc.gsfc.nasa.gov/docs/nicer/data_analysis/workshops/NICER-CalStatus-Markwardt-2021.pdf}}.

The best-fit parameters obtained from \nicer{} and \nustar{} broadband spectral fitting are -- photon index of $\sim$1.4, cutoff energy of $\sim$16 keV, the ionization parameter $\log{\xi} \sim$3.7 erg~cm~s$^{-1}$, and the iron abundance $\rm A_{Fe}$ $\sim 1$. The inner disk radius R$_{in}$, and disk inclination $i$ are found to be $9.5^{+0.5}_{-0.7} R_g$ and $50.3^{+2.0}_{-1.3}~^\circ$, respectively. Figure~\ref{fig:reflection_nustar_nicer} represents the best-fit broadband burst-free spectra of \obj{} with \nicer{} and \nustar{} using the reflection model. We summarize the best-fit spectral parameters in Table~\ref{tab:fitstat2}. The total unabsorbed flux in the energy band of 1-100 keV is estimated to be $3.5\times 10^{-9}$ ergs~cm$^{-2}$~s$^{-1}$ with corresponding luminosity of $4.2\times10^{37}$ ergs s$^{-1}$ assuming a source distance of 10 kpc. For a canonical neutron star with mass 1.4 $M_\odot$ and radius of 10 km, the Eddington luminosity ($L_{\textrm{Edd}}$) is $\sim3.8\times10^{38}$ ergs s$^{-1}$ \citep{Ku03}. Therefore, the estimated persistent luminosity of the source corresponds to 11\% of the Eddington luminosity. The flux of the reflection component is estimated to be $3.2\times 10^{-9}$ ergs~cm$^{-2}$~s$^{-1}$. The Markov Chain Monte Carlo {\tt MCMC} technique is used to estimate errors on the spectral parameters, and Figure~\ref{fig:corner} presents the {\tt MCMC} chain corner plot. The Goodman-Weare algorithm \citep{Good2010} is employed to conduct the {\tt MCMC} simulation, using 10 walkers and a chain length of 300,000. Assuming that the first 30,000 steps were in the `burn-in' or `transient' phase, we opted out.

The light curve of the source (Figure~\ref{fig:MAXI}) shows a significant variation in the flux during the X-ray outburst. As the source was observed with \nicer, \xmm, and \nustar~ at different flux levels, it is worth investigating the evolution of spectral parameters in the burst-free persistent emission. As discussed above, the persistent emission is characterized by the relativistic disk reflection model. As \nicer{} observations are limited up to 10 keV, and the reflection features are prominent beyond 10 keV (Compton hump), \nicer~ data alone may not be adequate to model this feature. Due to the cross calibration uncertainties and presence of several known instrumental features at low energies in the \xmm{} EPIC-PN data, and the presence of low-energy residuals due to the bright objects that are observed in timing mode \citep{Da10}, it may not be suitable to use along with the \nustar~ data to characterize the persistent emission at different flux levels. This may affect the spectral fitting results. Considering this, we selected two segments from the \nustar~ data, as marked in Figure~\ref{fig:burst_combined}, to carry out the study of the evolution of persistent spectral parameters at different flux levels. The \nustar{} spectra at these two different segments are fitted with {\tt TBabs $\times$ (relxill + bbodyrad)}. We used the \nustar{} 3-50 keV spectra to investigate the evolution of spectral parameters in two burst-free segments at different flux levels. The best-fit spectral parameters are summarized in Table~\ref{tab:fitstat2}. The inner disk radii for \nustar{} Seg-1 and Seg-2 are found to be $\sim11~R_g$ and $\sim~12 R_g$, respectively. The inclination angle is found to be $53.0\pm2.0^\circ$  and $46\pm 2.5^\circ$ for \nustar{} Seg-1 and Seg-2, respectively. The total unabsorbed flux during Seg-1 is estimated to be $\sim3.9\times10^{-9}$ erg cm$^{-2}$ s$^{-1}$, which is approximately 45\% higher compared to that during the Seg-2. The reflection fraction is found to be higher during \nustar{} Seg-1, and the flux of the reflection component is $\sim3.5\times10^{-9}$ erg cm$^{-2}$ s$^{-1}$, which is $\sim$50\% higher compared to the flux of the reflection component during \nustar{} Seg-2. This indicates that at higher flux levels, the reflection component contributes more towards the total emission. 

\begin{table*}
	\centering
    \normalsize
	\caption{Fitting parameters for the broadband persistent spectra of \obj{}. The spectral model M1 yields the best fit, M1: \texttt{tbabs $\times$ (bbodyrad + relxill)}. All errors are 90\% significant and are computed using the MCMC in {\tt XSPEC}.}
	\label{tab:fitstat2}

\begin{tabular}{|l|c|c|c|c|}
	\hline
	Parameters & \nicer{}+\nustar{} & \nustar{} Seg-1 & \nustar{} Seg-2  \\
                   & (Simultaneous)     &                 &  \\
	\hline
	N$_\mathrm{H}$ ($10^{22}$ cm$^{-2}$) & $2.5 \pm 0.1$ & $3^f$ & $3^f$   \\

        $kT_{\rm bb}$ (keV) & $0.60 \pm 0.01$ &$0.56 \pm 0.01$ & $0.50\pm 0.01$  \\

        Norm                & $229^{+15}_{-20}$             &  $437\pm15$ & $801\pm80$  \\
 
	 Incl (deg) 	&	 $50.3_{-1.3}^{+2.0}$ 	&	$53.0\pm 2.0$ & $46\pm 2.5$  \\
        	
	 R$_{\textrm{in}}$ (R$_{\textrm{ISCO}})$	& $1.8_{-0.4}^{+0.3}$ 	&	$2.1\pm0.5$ 	& $2.2\pm 0.6$  \\
 
 	 $E_{\textrm{cut}}$ (keV) 	&	 $16.0\pm1.0$ 	&	$20.0\pm 0.2$ & $21.0\pm 0.5$  \\

$\Gamma$	&	 $1.41\pm0.05$ 	&  $1.52\pm 0.01$ & $1.6\pm 0.02$  \\

	 $\rm A_{Fe}$ 	&	 $1.0\pm0.1$ 	&	$0.6\pm 0.1$ & $1.1\pm 0.2$   \\

	 q	&	 $3^f$ 	&	 $3^f$ & $3^f$  \\
 	 $a^\ast$	&	$0.21^{f}$ &	$0.21^f$ 	& $0.21^f$   
   \\

  $\log{\xi}$ (erg cm s$^{-1}$) 	&	 $3.70\pm0.04$ 	&	 $3.70\pm 0.05$ & $3.70\pm 0.05$  \\

  Refl$_{\textrm{frac}} $ 	&	 $0.70\pm 0.15$ 	&	$2.1\pm 0.1$ & $0.45\pm 0.05$  \\

  norm ($\times10^{-3}) $	& $3.1_{-0.1}^{+0.2}$ 	&	$1.9 \pm 0.1$  & $2.6\pm 0.2$  \\
 \hline
	 Reflection Flux$^a$ 	&	 $3.24\pm0.01$ 	&	$ 3.46\pm 0.01$  & $2.30\pm0.01$  \\

      Total Flux$_{\textrm{1-100~keV}}^b$ 	&	 $3.48\pm0.01$ 	&	$ 3.88 \pm 0.01$  & $2.66\pm0.01$  \\
                   	
	 Luminosity$_{\textrm{1-100~keV}}^c$ 	&	 $4.18 \pm 0.01$ 	& $ 4.64 \pm 0.01$ & $3.18\pm0.01$  \\

\hline                   																
	 $\chi^2$/dof           	& 2580/2557 &	   1867/1763    & 1819/1769       	\\
\hline					
    \multicolumn{4}{l}{$^a$ : in the units of $10^{-9}$ \erg.}\\
    \multicolumn{4}{l}{$^b$ : Total unabsorbed flux in the units of $10^{-9}$ \erg.}\\
    \multicolumn{4}{l}{$^c$ : X-ray luminosity in the units of $10^{37}$ \lum, assuming a source distance of 10 kpc.}\\
    \multicolumn{4}{l}{$^f$ : Frozen parameters.}\\

	\end{tabular}
    
\end{table*} 

\section{Discussion}
\label{dis}

We report the results from the detailed time-resolved spectral study of thermonuclear bursts using \nicer{} to probe the dynamic evolution of different spectral parameters. We also performed a time-resolved spectral study of a burst simultaneously detected with \xmm{} and \nustar{}. The energy dependence of the burst profiles is investigated using \nicer{}, \xmm{} EPIC-PN, and \nustar{} data, which indicates a strong dependence of the burst profile on energy. We do not find compelling evidence for photospheric radius expansion (PRE) in the analyzed bursts. However, the use of comparatively broad time segments may limit sensitivity to short-lived expansion episodes. The burst spectra can be modeled with an absorbed blackbody component. The results of time-resolved spectral analysis using the absorbed blackbody model are comparable with the earlier findings \citep{Ng24, Malacaria2025}. In addition, we investigated the reflection features during the \nicer{} bursts with a comparatively higher time bin (as discussed in Section~\ref{reflection_burst}), which indicates a deviation of burst emission from blackbody emission. Several possible explanations exist behind this, including the atmospheric effect \citep{Ozel2013} and Poynting–Robertson drag \citep{Walker1992, Worpel2013}, among others. Apart from this, a fraction of the burst photons may interact with the disk and be reflected from the accretion disk. We find that the burst spectra can be explained with a more realistic disk reflection model {\tt relxillNS}, which provides a physically motivated and self-consistent explanation. This approach was adopted to model the time-resolved burst spectra for Aql~X-1 \citep{Ma25}, 4U~1730--22 \citep{Lu23}, 4U~1820--30 \citep{Ke18a, Ja24}, 4U~1636--536 \citep{Zh22}, 4U~1702-729 \citep{4U1702}, and SAX~1808.4-3658 \citep{Bu21}. It is found that the reflection component can contribute up to $\sim$30\% of the total burst emission as observed during the peak of \nicer{} bursts. The flux of the reflection component is found to be positively correlated with the blackbody flux. 

\subsubsection{Bursting regime}
Depending on the mass accretion rate, the thermonuclear ignition regime can be probed. The local accretion rate onto the NS can be expressed as \citet{Ga08} 

\begin{align}
\dot{m} =\; & \frac{L (1+z)}{4\pi R^2 (GM/R)} \nonumber \\
=\; & 6.7 \times 10^{3} 
\left( \frac{F_{b}}{10^{-9}~\mathrm{erg~cm^{-2}~s^{-1}}} \right) 
\left( \frac{d}{10~\mathrm{kpc}} \right)^2 
\left( \frac{1+z}{1.31} \right) \nonumber \\
& \times \left( \frac{R}{10~\mathrm{km}} \right)^{-1} 
\left( \frac{M}{1.4~M_\odot} \right)^{-1} 
~\mathrm{g~cm^{-2}~s^{-1}}
\end{align}

where $F_b$ is the persistent bolometric flux of \obj{} (Table~\ref{tab:tab_preburst}), $d$ is the distance to the NS, and $M$ and $R$ are the mass and radius of the NS, respectively. Assuming a typical NS of radius 10 km, mass of 1.4$~M_\odot$, gravitational redshift ($z$) of 0.3, and source distance of 10 kpc, we estimated the mass accretion rate for persistent flux before the \nicer{} TNB-2 as $5.8~\times~10^{4}$ gm~cm$^{-2}$~s$^{-1}$ corresponding to $F_b$ = $8.6~\times~10^{-9}$ erg~cm$^{-2}$~s$^{-1}$. The details related to the mass accretion rate are summarized in Table~\ref{tab:tab_preburst}. Now, the local Eddington mass accretion rate for a typical NS with radius 10 km is $\dot{m}_{\mathrm{Edd}} = 8.8~\times~10^4~\mathrm{g~cm^{-2}~s^{-1}}$. The local accretion rate is related to the Eddington rate as $\dot{m} = 0.6~\dot{m}_{\mathrm{Edd}}$, which is higher than the required rate for stable burning of hydrogen in helium. The estimated mass accretion rate indicates that, at the observed values of local mass accretion rates, it is unlikely to have a PRE burst. The higher accretion rate causes helium to ignite unstably in an H-rich environment, since hydrogen is accreted more quickly than it can be burned by steady burning \citep{Ga08}. The measurements on the mass accretion rate indicate that the burst fuel composition is possibly mixed H/He. Typically, for a mixed H/He powered burst, the burst shows a duration of 10--100 s, ignition column depth of $10^8$ gm~cm$^{-2}$, and burst fluence of $10^{39}$ erg \citep{Ga21}. Earlier measurements on the $\alpha$ factor and the depth of the ignition column indicate a possible mixed H/He-powered burst \citep{Malacaria2025}. However, a discrepancy was observed between the observed and estimated recurrence times, which may hint at a peculiar burning fuel composition or, assuming that the burst occurred in a pure He environment, the observed and estimated recurrence times closely match \citep{Malacaria2025}.
 
\subsubsection{Reflection from persistent emission: Estimation of magnetic field}

The observed reflection features in persistent broadband spectra can be well modeled using the relativistic reflection model \relxill{} along with a blackbody model component. Modern and high-resolution telescopes such as \xmm{}, \nustar{}, and {\it INTEGRAL} have been used to observe X-ray reflection in various AMXPs \citep{Pa10, Pa13a, Pa13b, Pa16, 4U1702, Ma25, Lu19, Ch22}. In addition to the reasonably broad iron line, the inner disk radius is observed to be quite small in different AMXPs \citep{Pa10, Pa16}. The $\rm R_{ISCO}$ for a rotating neutron star with the spin parameters $a^\ast$ = 0.21 can be defined as \citet{Va04}, 

\begin{align}
R_{\rm ISCO} & = \frac{6GM}{c^2} \left(1 - 0.54 a^\ast \right) 
= 5.3 \, \frac{GM}{c^2} = 5.3\, R_g
\end{align}

where, $M$ is the mass of the neutron star, $G$ is the gravitational constant, and $c$ is the speed of light. The inner disk radius is estimated from the \nicer{} and \nustar{} best-fit spectral parameters, $\rm R_{in}\sim 2.0~R_{ISCO} \sim10.6 R_g$. Typically, the inner disk radius in AMXPs is below 15 $R_g$ \citep{Pa09}; however, a larger value ($\sim40 R_g$) was also observed in different sources \citep{Pa10, Pa13a}.

Due to the pressure of the magnetic field, the accretion disk of NS LMXB is expected to be truncated at a moderate radius \citep{Ca09}. It is possible to determine the magnetic field strength, assuming the disk is truncated at the magnetospheric radius. The magnetic dipole moment ($\mu$) and the magnetic field can be estimated by using the equation from \citet{Ib09}, 

\begin{align}
\mu =\; & 3.5 \times 10^{23} \, x^{7/4} \, k_A^{-7/4} 
\left( \frac{M}{1.4 M_\odot} \right)^2 
\left( \frac{f_{\rm ang}}{\eta} \frac{F_b}{10^{-9}~\rm erg~cm^{-2}~s^{-1}} \right)^{1/2} \nonumber \\
& \times \left( \frac{D}{3.5~\rm kpc} \right)
\end{align}
where, $\eta$ is accretion efficiency, $\rm f_{ang}$ is anisotropy correction factor and $k_A$ is geometric coefficient, $M$ is the mass of the NS, and $F_b$ is the unabsorbed bolometric flux in 1--100 keV range. The scaling factor $x$ can be estimated from $\text R_{\rm in} = \frac{xGM}{c^2}$. Assuming $\eta = 0.1$, $f_{\rm ang} = 1$, and $k_A = 1$ \citep{Ca09}, the magnetic dipole moment of the NS can be estimated to be $\mu = 3.0 \times 10^{26}$ G~cm$^3$ for the flux of $\text F_b \sim 3.5~\times~10^{-9}$ erg~cm$^{-2}$~s$^{-1}$. The magnetic field strength at the poles of NS of radius 10 km is estimated to be $ B~\sim 6.0~\times~10^8$ G. The estimated magnetic field is in the typical range of NS LMXBs \citep{Mu15, Ca09, Lu16}.

\section{Conclusion}
\label{con}
We conducted a broadband spectral and temporal study of \obj{} during thermonuclear bursts, detected at various phases of the outburst using \nicer{}, \xmm{}, and \nustar{} observations. The burst profiles exhibit a strong energy dependence, as observed with \nicer{}, \xmm{}, and \nustar{}, and are detected up to 20 keV. We reported the findings from the time-resolved spectral study for \nicer{} bursts and a simultaneously detected burst with \xmm{} and \nustar{}. We also investigated the reflection features during the bursts using the disk reflection model {\tt relxillNS}. The results suggested that the reflection component can contribute up to $\sim$30\% of the total burst emission. A positive correlation is reported between the reflection and blackbody component fluxes. The ratio of the pre-burst mass accretion rate to the Eddington rate is found to lie between $\sim$0.4-0.7. The estimated mass accretion rate indicates that the burst may be powered by mixed H/He fuel, and a higher accretion rate may cause helium to ignite unstably in an H-rich environment. The burst-free persistent broadband \nicer{}, and \nustar{} spectra show the presence of a reflection feature, which is modeled with the relativistic disk reflection model {\tt relxill}. The inner disk radius and disk inclination angle are found to be $\sim 11R_g$ and $50.3^{+1.3}_{-2.0}~^\circ$, respectively. Based on the results, the magnetic field is estimated as $B~\sim 6.0 \times   10^8$ G, assuming the disk is truncated at the magnetospheric radius.

\section*{Acknowledgements}
We thank the referee for useful comments, which helped to improve the manuscript. The research work at the Physical Research Laboratory, Ahmedabad, is funded by the Department of Space, Government of India. The authors are thankful to Alessandro Papitto and Christian Malacaria for their comments on the draft. This research has made use of data obtained with \nustar{}, a project led by Caltech, funded by NASA, and managed by NASA/JPL, and has utilized the {\tt NUSTARDAS} software package, jointly developed by the ASDC (Italy) and Caltech (USA). We acknowledge the use of public data from the \nustar{}, \nicer{}, and \xmm{} data archives.

\section*{Data Availability}
The data used for this article are publicly available in the High Energy Astrophysics Science Archive Research Centre (HEASARC) at \\
\url{https://heasarc.gsfc.nasa.gov/db-perl/W3Browse/w3browse.pl}.


\bibliographystyle{mnras}
\bibliography{SRGA_MNRAS}



\appendix

\section{Burst profile and energy dependence}

\begin{figure*}
\centering{
\includegraphics[width=4cm]{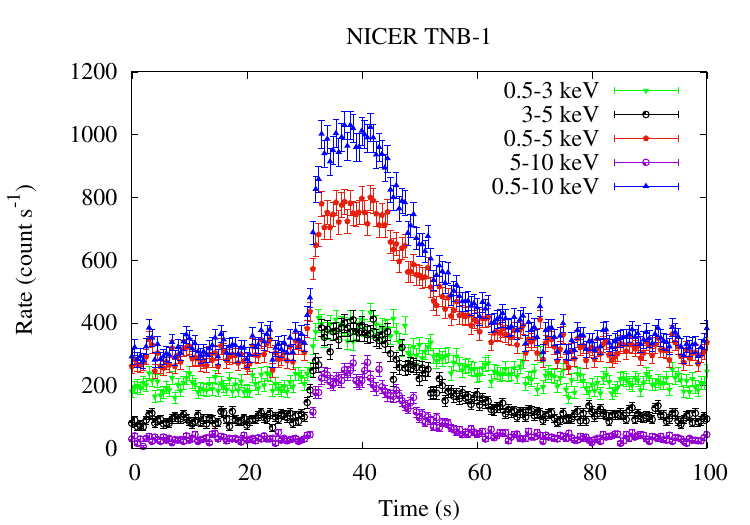}
\includegraphics[width=4cm]{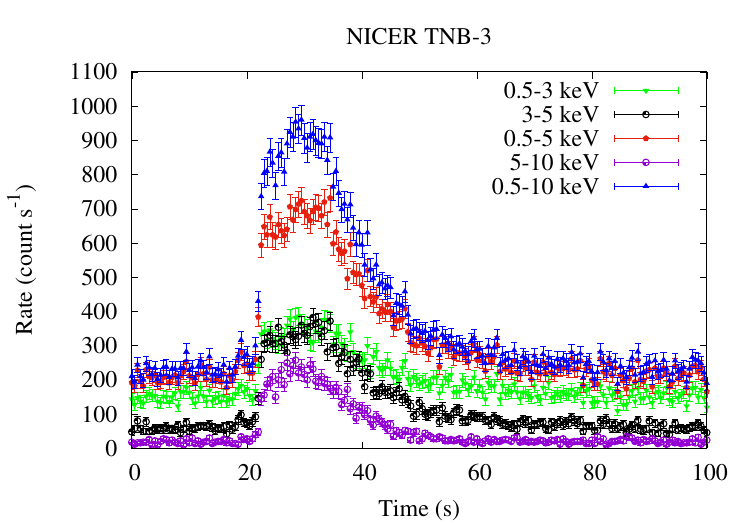}
\includegraphics[width=4cm]{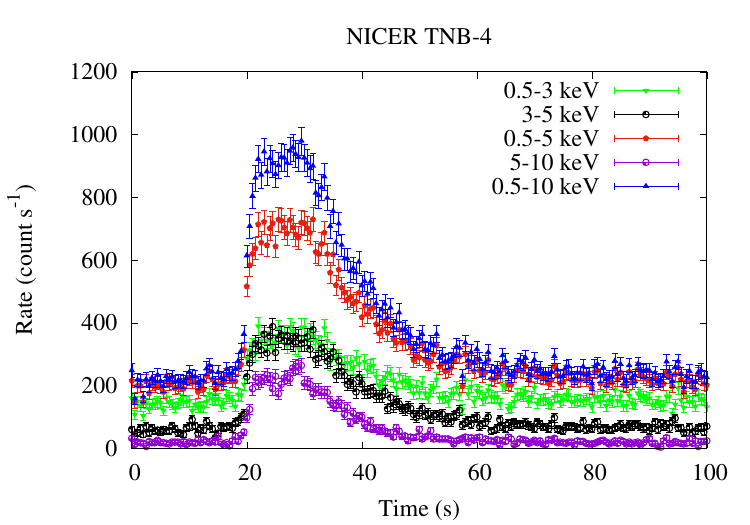}
\includegraphics[width=4cm]{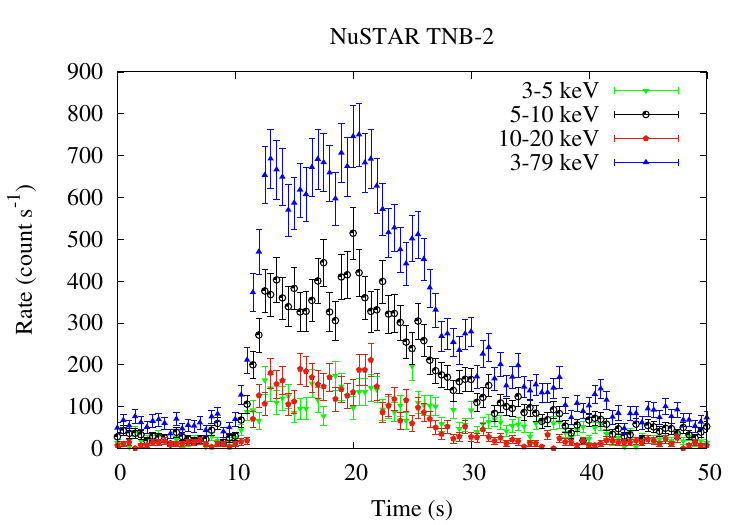}
\includegraphics[width=4cm]{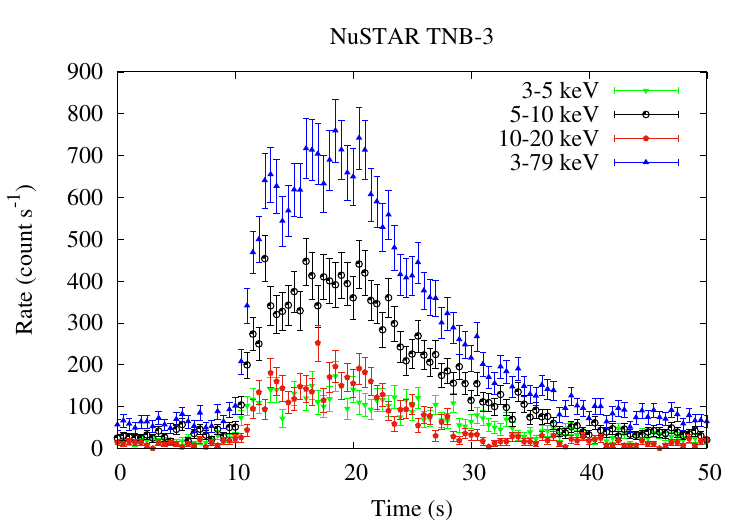}
\includegraphics[width=4cm]{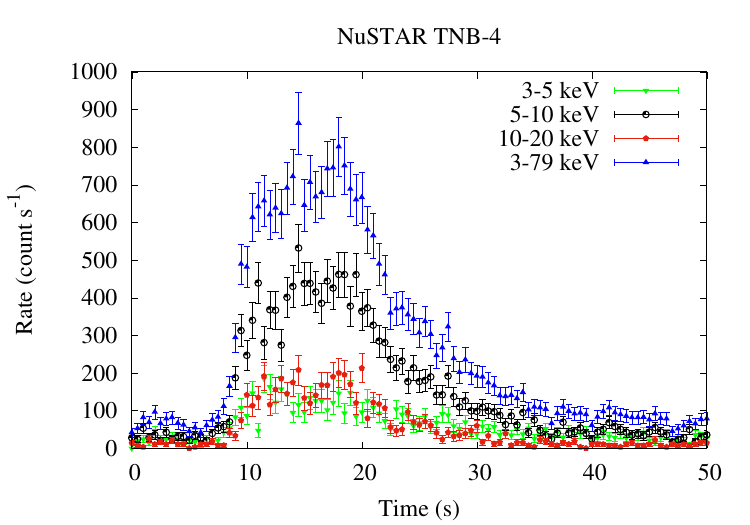}
\includegraphics[width=4cm]{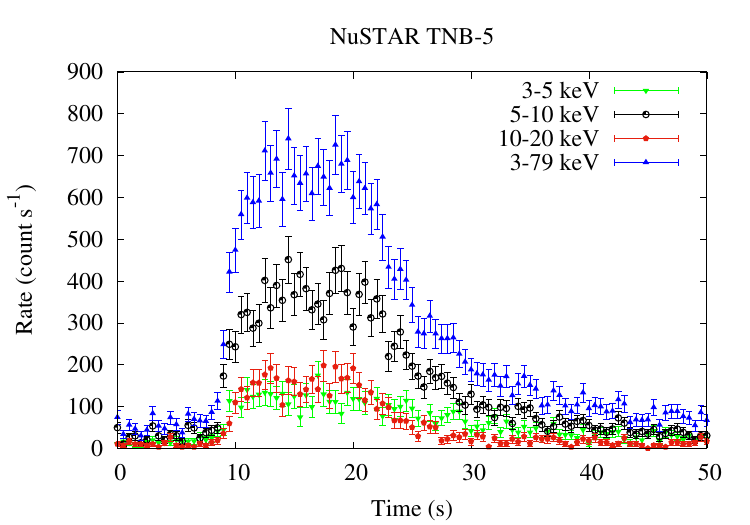}
\includegraphics[width=4cm]{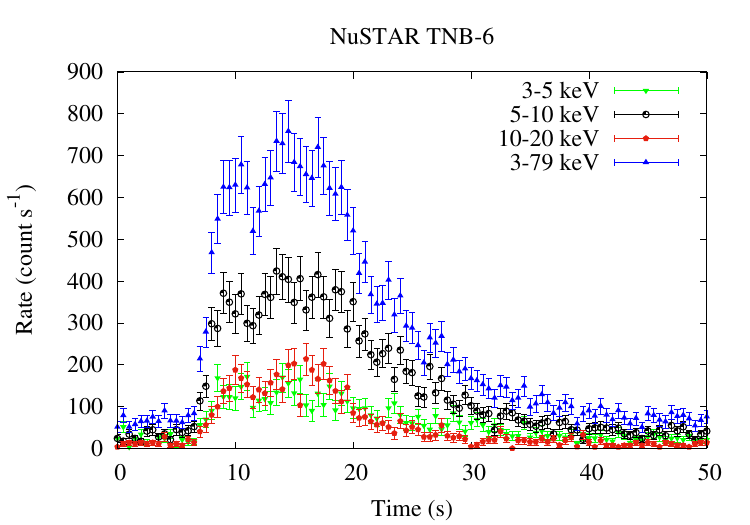}
\includegraphics[width=4cm]{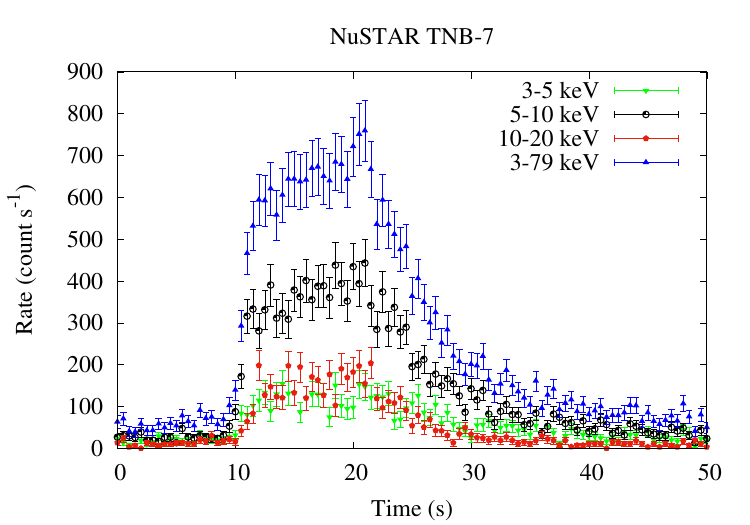}
\includegraphics[width=4cm]{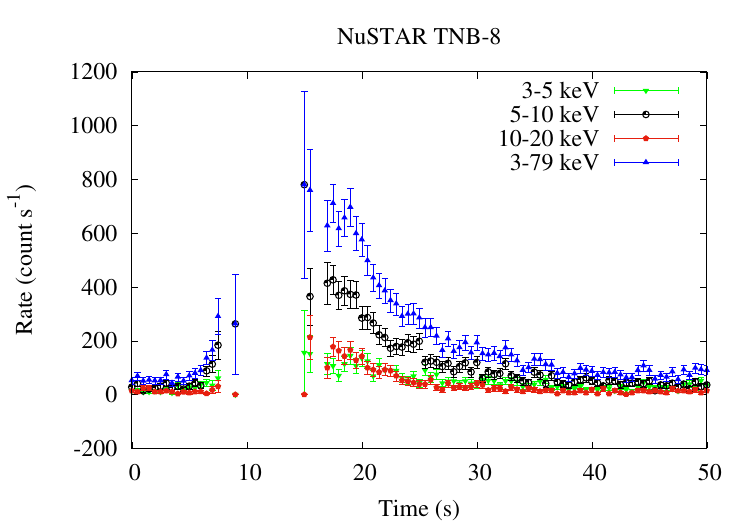}
\includegraphics[width=4cm]{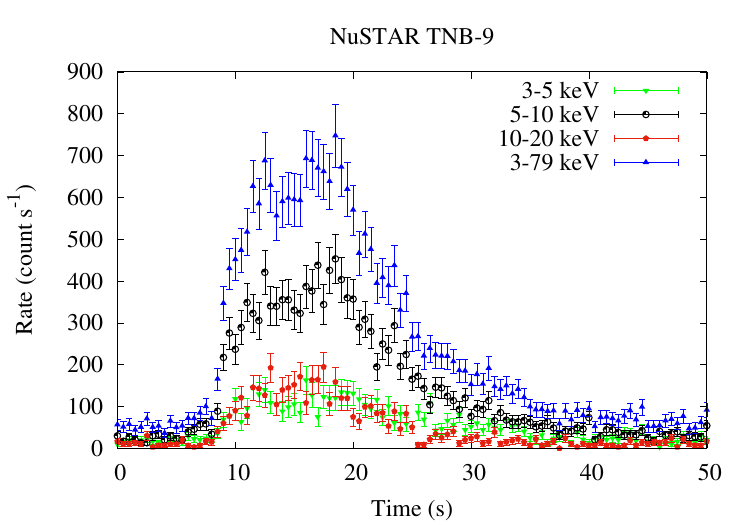}
\includegraphics[width=4cm]{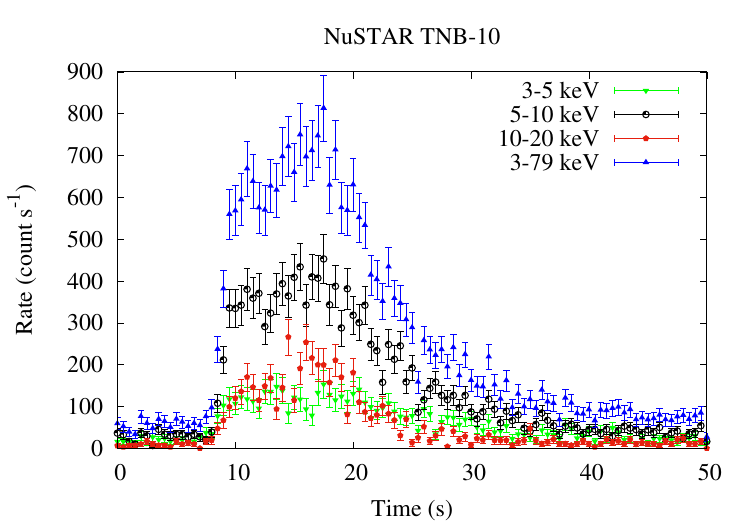}
\includegraphics[width=4cm]{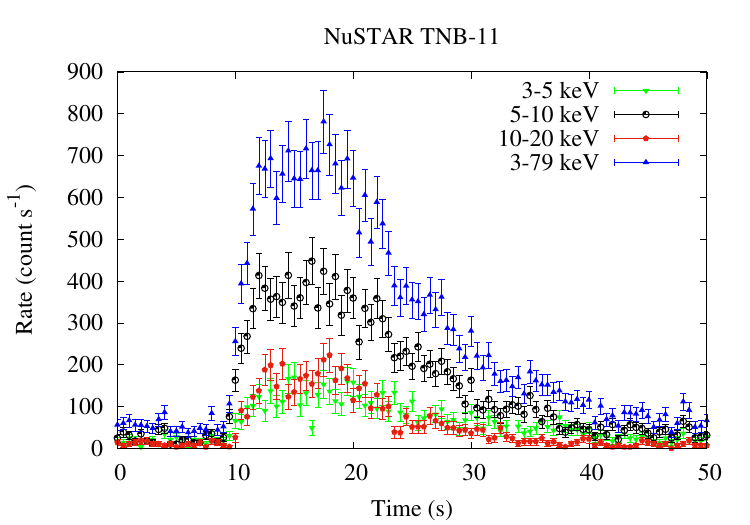}
\includegraphics[width=4cm]{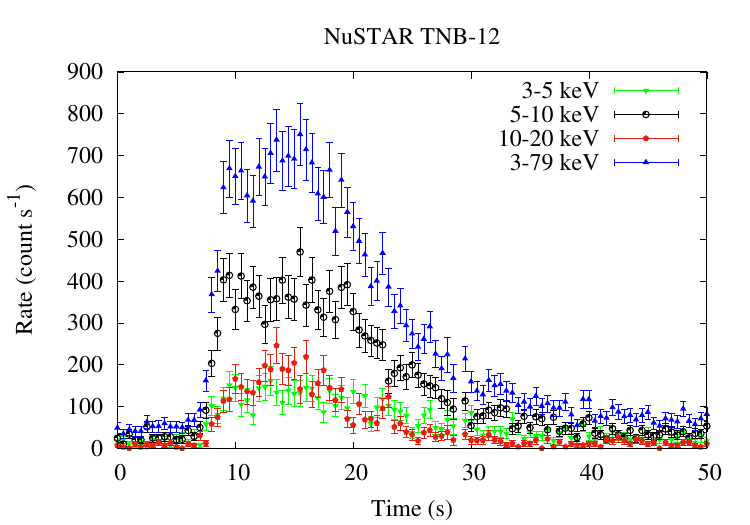}
\includegraphics[width=4cm]{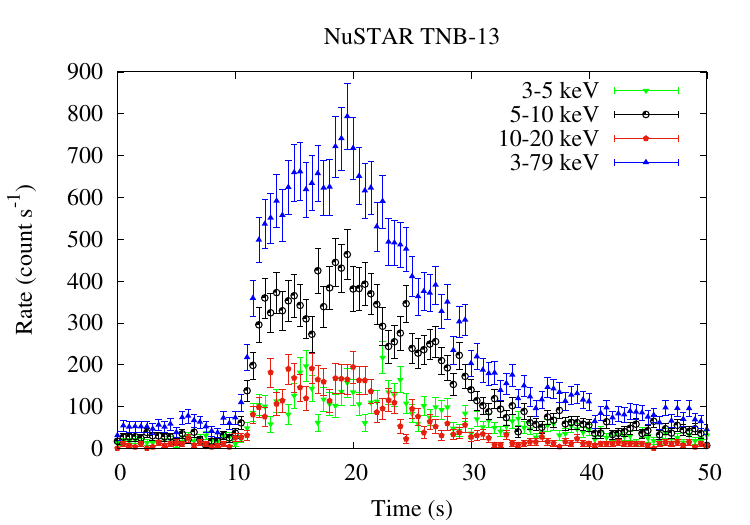}
\includegraphics[width=4cm]{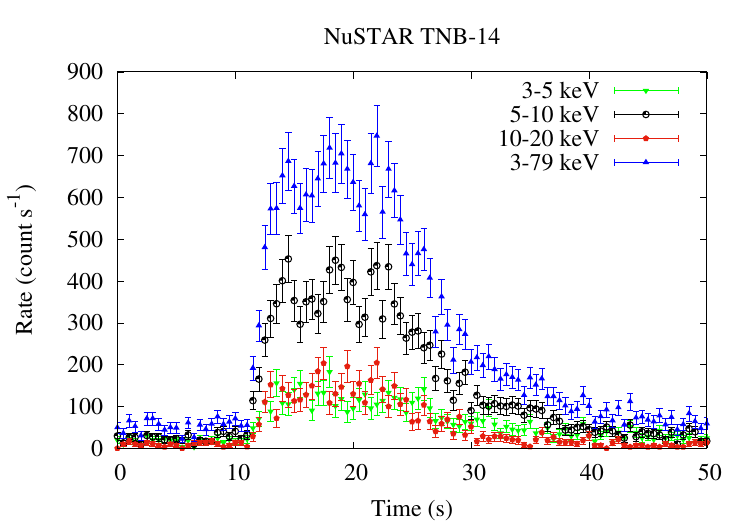}
\includegraphics[width=4cm]{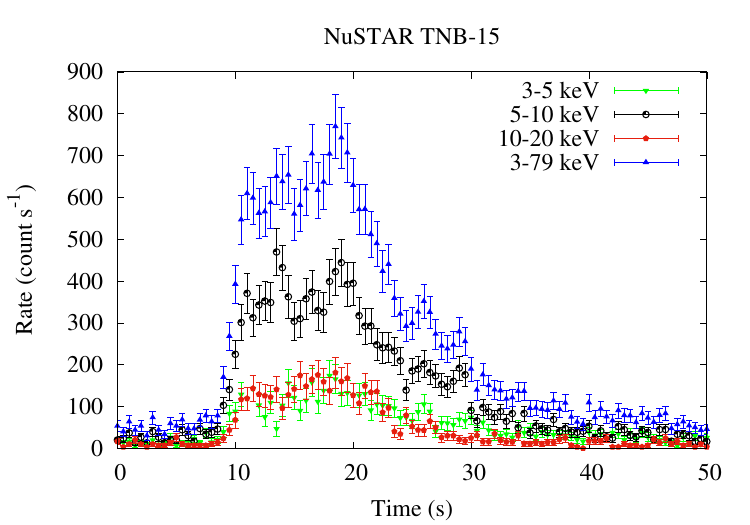}
\includegraphics[width=4cm]{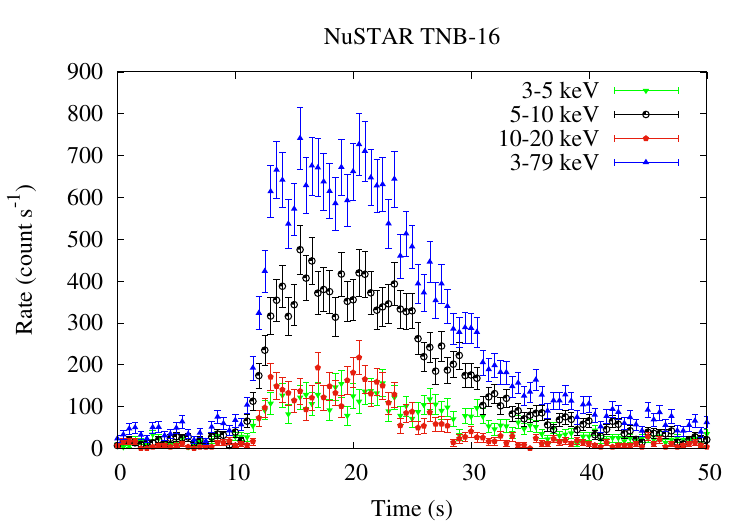}
\includegraphics[width=4cm]{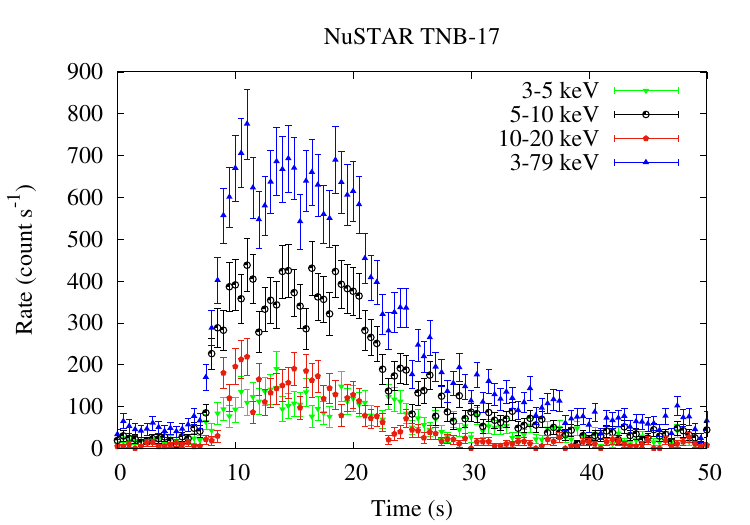}
\includegraphics[width=4cm]{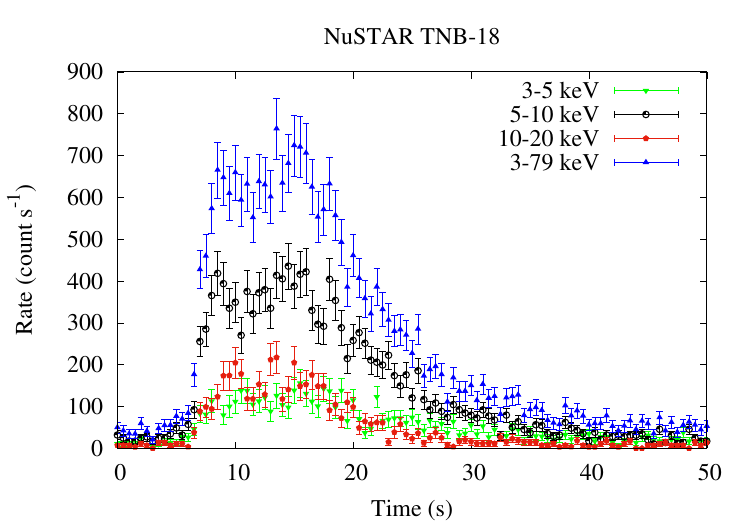}
\includegraphics[width=4cm]{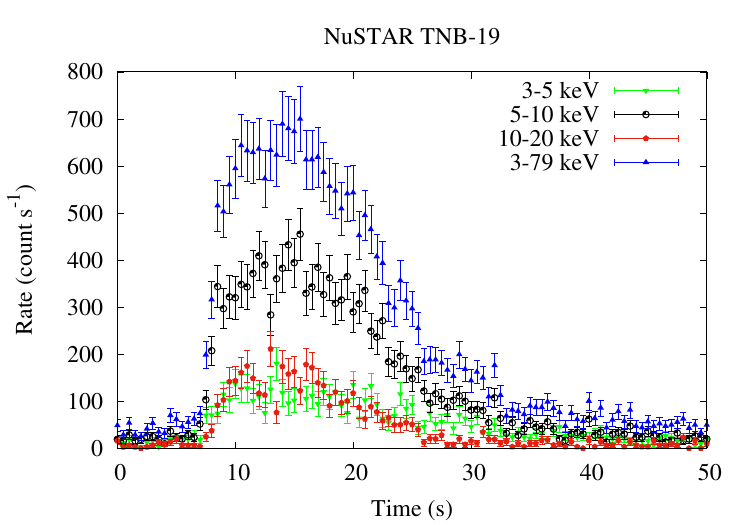}
\includegraphics[width=4cm]{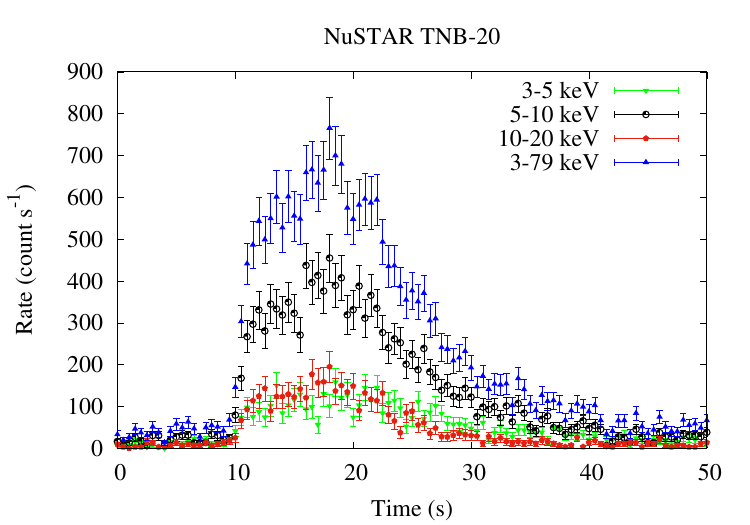}
\includegraphics[width=4cm]{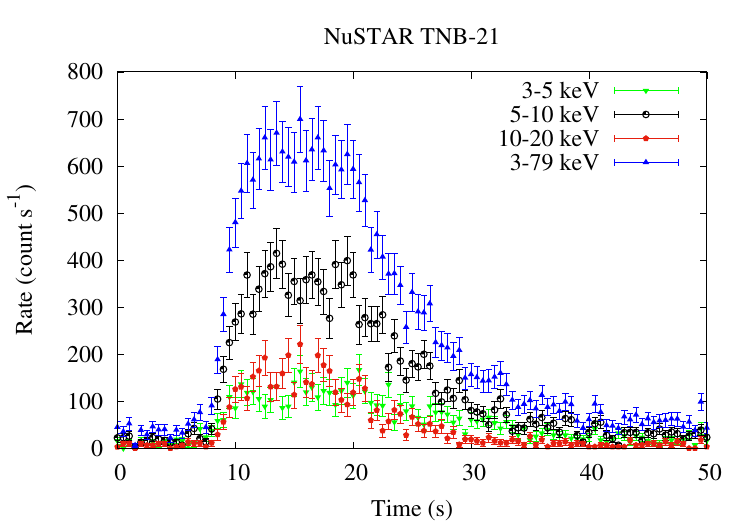}
\includegraphics[width=4cm]{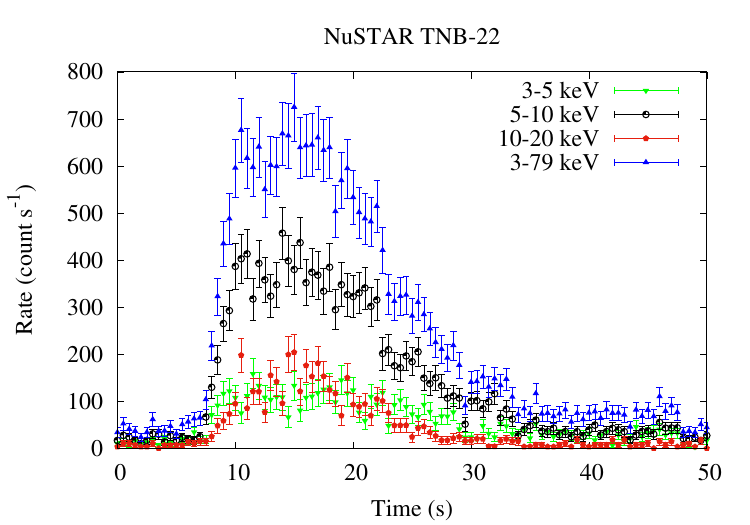}
\caption{Energy-resolved burst profiles of \src{} using \nicer{} and \nustar{}.}
	 \label{fig:burst_profile_all}}
\end{figure*}

Figure~\ref{fig:burst_profile_all} shows the energy dependence of the burst profiles for different thermonuclear X-ray bursts, which were detected from \obj{} with \nicer{} and \nustar{}.

\section{MCMC corner plot}
\begin{figure*}
\centering
 \includegraphics[width=\linewidth]{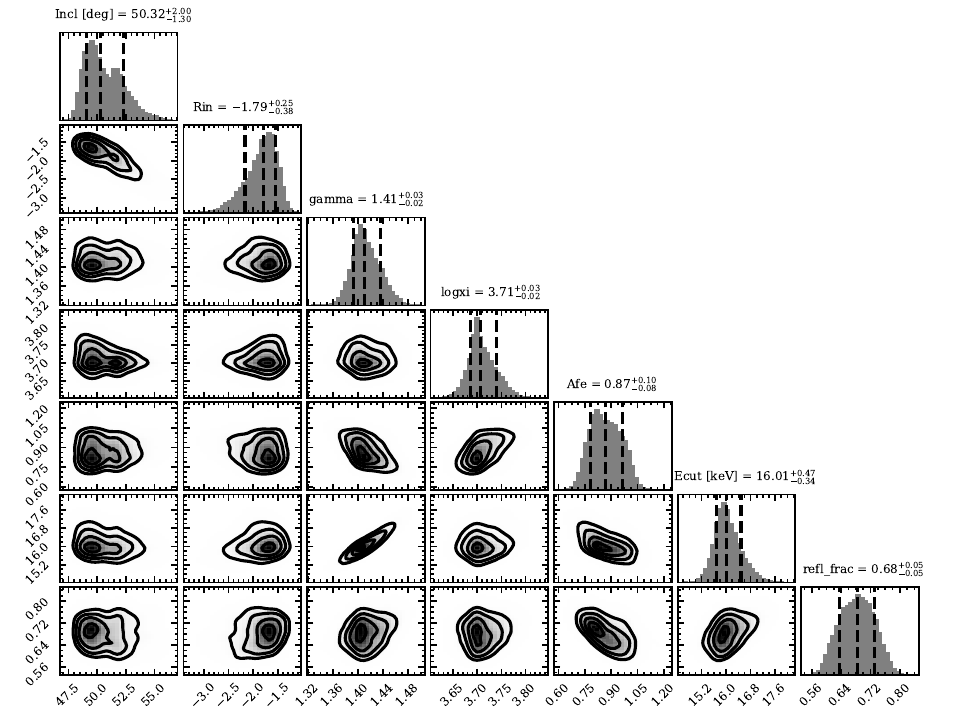}
 \caption{MCMC chain corner plot is shown for the broadband \nicer{} and \nustar{} spectral modeling of \src{} with the best-fitting model M4.}
\label{fig:corner}
\end{figure*}

Figure~\ref{fig:corner} shows the MCMC corner plot for model M4: \texttt{tbabs$\times$(relxill+bbodyrad)}.

\pagestyle{plain}

\bsp	
\label{lastpage}
\end{document}